\documentclass[fleqn,10pt]{wlscirep}
\usepackage[utf8]{inputenc}
\usepackage[T1]{fontenc}
\usepackage{bm}

\usepackage{graphicx}
\usepackage{dcolumn}
\usepackage{lmodern}%
\usepackage{hyperref}%
\usepackage{xcolor}%
\usepackage{placeins}
\usepackage{tabularx}
\usepackage{booktabs}
\usepackage[normalem]{ulem}
\usepackage{placeins}

\usepackage[justification=justified]{caption}

\newcommand{\beq}{\begin{equation}\begin{aligned}}
\newcommand{\eeq}{\end{aligned}\end{equation}}

\newcommand{\sio}{SiO\ensuremath{_2}}

\newcommand{\ws}{WS\ensuremath{_2}}
\newcommand{\wse}{WSe\ensuremath{_2}}
\newcommand{\mos}{MoS\ensuremath{_2}}
\newcommand{\mose}{MoSe\ensuremath{_2}}
\newcommand{\mote}{MoTe\ensuremath{_2}}

\newcommand{\vthe}{\ensuremath{V_{th}^e}}
\newcommand{\vthh}{\ensuremath{V_{th}^h}}
\newcommand{\isd}{\ensuremath{I_{SD}}}
\newcommand{\vsd}{\ensuremath{V_{SD}}}
\newcommand{\vref}{\ensuremath{V_{\rm Ref}}}
\newcommand{\figref}[2]{\ref{#1}\textsf{\bfseries #2}}

\newcommand{\ie}{\emph{i.e.},}
\newcommand{\eg}{\emph{e.g.}}

\newcommand{\dqmp}{Department of Quantum Matter Physics, University of Geneva, 24 Quai Ernest Ansermet, CH-1211 Geneva, Switzerland}
\newcommand{\gap}{Department of Applied Physics, University of Geneva, 24 Quai Ernest Ansermet, CH-1211 Geneva, Switzerland}

\definecolor{linkcol}{rgb}{0,0,0.4}
\definecolor{citecol}{rgb}{0.5,0,0}

\hypersetup{
	bookmarksopen=true,
	pdftitle="Ionic Gate Spectroscopy of 2D Semiconductors",
	pdfauthor="IGL et al",
	pdfsubject="Ionic Gate Spectroscopy of 2D Semiconductors", 
	pdfstartview={FitH},    
	pdfmenubar=true, 
	pdfhighlight=/O, 
	colorlinks=true, 
	pdfpagemode=UseNone, 
	pdfpagelayout=SinglePage, 
	pdffitwindow=true, 
	linkcolor=linkcol, 
	citecolor=linkcol, 
	urlcolor=blue
}

\title{Ionic Gate Spectroscopy of 2D Semiconductors}

\author[1,2]{Ignacio Gutiérrez-Lezama}
\author[1,2]{Nicolas Ubrig}
\author[1,2]{Evgeniy Ponomarev}
\author[1,2]{Alberto F. Morpurgo}
\affil[1]{\dqmp}
\affil[2]{\gap}

\affil[*]{e-mail: alberto.morpurgo@unige.ch}

\begin{abstract}
	Reliable and precise measurements of the relative energy of band edges in semiconductors are needed to determine band gaps and band offsets, as well as to establish the band diagram of devices and heterostructures. These measurements are particularly important in the field of two-dimensional materials, in which many new semiconducting systems are becoming available through exfoliation of bulk crystals. For two-dimensional semiconductors, however, commonly employed techniques suffer from difficulties rooted either in the physics of these systems, or of technical nature. The very large exciton binding energy, for instance, prevents the band gap to be determined from a simple spectral analysis of photoluminescence, and the limited lateral size of atomically thin crystals makes the use of conventional scanning tunneling spectroscopy cumbersome. Ionic gate spectroscopy is a newly developed technique that exploits ionic gate field-effect transistors to determine quantitatively the relative alignment of band edges of two-dimensional semiconductors in a straightforward way, directly from transport measurements (\ie\ from the transistor electrical characteristics). The technique relies on the extremely large geometrical capacitance of ionic gated devices that --under suitable conditions-- enables a change in gate voltage to be directly related to a shift in chemical potential. Here we present an overview of ionic gate spectroscopy, and illustrate its relevance with applications to different two-dimensional semiconducting transition metal dichalcogenides and \emph{van der Waals} heterostructures.
\end{abstract}
\begin{document}

\flushbottom
\maketitle

\thispagestyle{empty}

\section*{Introduction}

The electrostatic accumulation of charge carriers at the surface of a semiconductor, at the core of field-effect transistors and of modern integrated electronics, is also instrumental in the investigation of the electronic properties of new materials\cite{ahn_electrostatic_2006}. In conventional transistor devices, electrostatic control of the charge density is achieved by a metallic gate electrode separated from the semiconductor by an insulating layer\cite{sze_physics_2006}. An alternative strategy to transfer the potential from the gate electrode to the surface of a semiconductor consists in using electrolytes with movable charged ions, \ie\ ionic gating (Fig.~\ref{fig:1})\cite{brattain_experiments_1955,bergveld_development_1970,kruger_electrochemical_2001,nilsson_bi-stable_2002,rosenblatt_high_2002,panzer_low-voltage_2005,shimotani_direct_2005,misra_electric_2007,fujimoto_electric-double-layer_2013,kim_electrolyte-gated_2013,leger_iontronics_2016}. The extremely large geometrical gate capacitance that can be reached in these ionically gated devices enables the accumulation of charge densities in excess of $5\times 10^{14}$~electrons/cm$^2$, causing the emergence of new physical phenomena\cite{bisri_endeavor_2017}. Possibly the best-known example is the occurrence of gate-induced superconductivity in semiconducting transition metal dichalcogenides (TMDs), such as \mos\ or \ws  \cite{ye_liquid-gated_2010,ye_superconducting_2012,jo_electrostatically_2015,shi_superconductivity_2015,biscaras_onset_2015,costanzo_gate-induced_2016,costanzo_tunnelling_2018,lu_full_2018,piatti_multi-valley_2018,kouno_superconductivity_2018,zeng_gate-induced_2018,piatti_ambipolar_2019}, a phenomenon that perfectly illustrates how ionic gated devices can be exploited to control the electronic state of materials.\\

The very large capacitance of ionic gated devices, however, has more to offer. Whenever the quantum capacitance\cite{luryi_quantum_1988,davies_physics_1997,ilani_measurement_2006,xia_measurement_2009,ihn_semiconductor_2010} of the gated material --or equivalently its density of states (DOS) at the chemical potential-- is sufficiently small, the electrostatic coupling between gate and channel is virtually ideal\cite{braga_quantitative_2012}. That is: a shift in gate potential causes the potential of the transistor channel to shift by the same amount. This strong electrostatic coupling implies the possibility to observe phenomena (\eg, ambipolar transport in semiconductors\cite{zhang_ambipolar_2012,braga_quantitative_2012,ubrig_scanning_2014,ovchinnikov_disorder_2016}) and to operate devices (\eg, light emitting transistors\cite{zhang_electrically_2014,jo_mono-_2014,ponomarev_ambipolar_2015,gutierrez-lezama_electroluminescence_2016}) under low bias conditions. In addition, as it has been realized only recently, it allows the implementation of a new form of spectroscopy, to determine quantitatively the relative position of energy levels in semiconductors from simple transistor measurements\cite{braga_quantitative_2012,jo_mono-_2014,lezama_surface_2014,ponomarev_ambipolar_2015,gutierrez-lezama_electroluminescence_2016,ponomarev_semiconducting_2018,saito_ambipolar_2015,ponomarev_semiconducting_2018,ciarrocchi_thickness-modulated_2018,zhang_band_2019,reddy_synthetic_2020,piatti_orientation-dependent_2020}.\\

It is easy to understand how spectroscopic information can be extracted from ionic gated devices with ideal electrostatic coupling. Transistor measurements allow identifying the gate voltage values for which the chemical potential is located at characteristic energies: the threshold voltage for electron or hole conduction, for instance, corresponds to having the chemical potential at the edge of the conduction or valence band, respectively. Because for ideal electrostatic coupling a shift in gate voltage induces an identical shift in chemical potential, the band gap of a semiconductor can then be extracted directly from threshold voltages for electron and hole transport. Although these considerations may seem too simplistic to form the basis of a reliable spectroscopic technique, over the last years extensive tests performed on a variety of semiconductors have established that reproducible and accurate band gap values can be obtained. Our aim here is to review the technique --which is not yet broadly known-- with a specific focus on the domain of two-dimensional materials.\\

\section*{Ionic gating and its development}

\begin{figure}
	\centering
	\includegraphics[width=.9\textwidth]{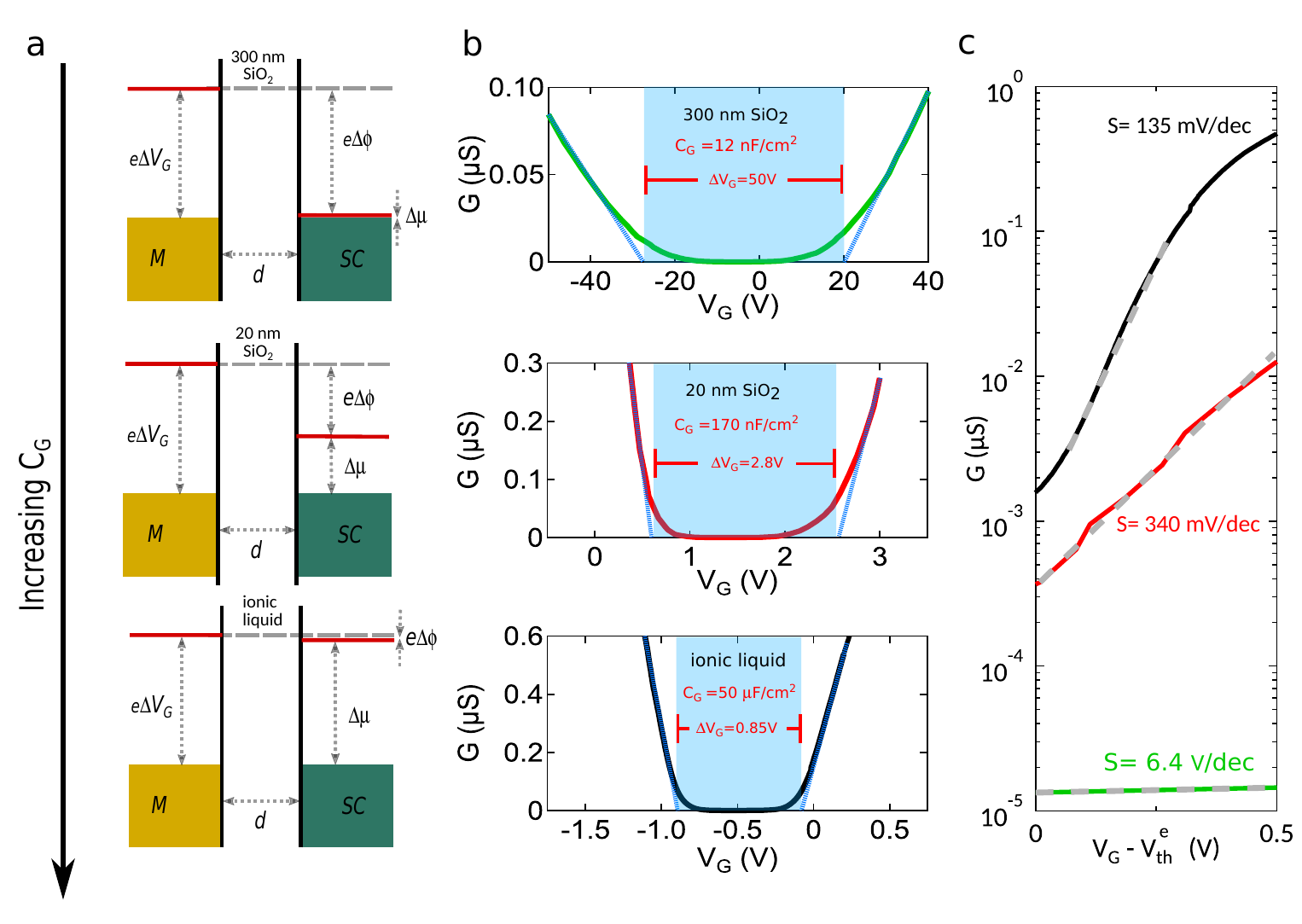}
	\caption{{\bfseries Evolution of the transistor characteristics upon increasing gate capacitance.}\footnotesize {\bfseries a.} In a field-effect transistor, a change in applied gate voltage $\Delta V_G$ causes a change in electrostatic potential difference ($\Delta \Phi$) and in chemical potential inside the semiconductor ($\Delta\mu$). The relative values of the gate geometrical capacitance  ($C_G$) and of the so-called quantum capacitance $C_Q$ (proportional to the density of states) determine the dominating contribution . {\bfseries b.} Transfer curves ($G$-vs-$V_G$) of transistors based  on thick exfoliated \mote\ crystals, with increasing  $C_G$, corresponding to the cases illustrated in {\bfseries a.} ($G$ is the square conductance or, equivalently for a 2D electronic system, the conductivity). Top panel:  device with  a 300 nm thick \sio\ layer as gate insulator ($C_G$~=~12~nF/cm$^2$; figure adapted from Ref.~\cite{yin_ferroelectric-induced_2017}). Shifting the chemical potential across the gap requires a change  $\Delta V_G \approx$~50~V, much larger than the MoTe$_2$ gap ($\Delta_{MoTe2}\simeq 0.85$ eV), showing that the  electrostatic potential dominates  ($\Delta V_G \approx \Delta\Phi$). Mid panel: for a  20~nm thick gate insulator ($C_G$~=~170~nF/cm$^2$), shifting the chemical potential across the gap requires only $\Delta V_G$~=~2.8~V, larger than but comparable to the band gap (figure adapted from Ref.~\cite{seo_low-frequency_2019}). Neither  $\Delta V_G$ nor $ \Delta\mu$ can be neglected. Bottom panel: the much larger capacitance of an ionic gate ($C_G \approx 50~ \mu$F/cm$^2$) makes $\Delta\Phi$ negligible,   and $e\Delta V_G\simeq \Delta\mu$. Indeed, $e(\vthe - \vthh)$~=~0.85~eV matches the MoTe$_2$ gap  (figure adapted from Ref.~\cite{lezama_surface_2014}). {\bfseries c.} From the slope  of the  conductance plotted in logarithmic scale we obtain the subthreshold slope $S$ ($S = \frac{kT}{e} \ln (10) (1+C_Q/C_G)$; the dashed lines are linear fits from which the value of $S$ is extracted). $S$ is found to decrease upon increasing $C_G$ and to approach the ultimate room-temperature limit of 60 mV/decade in ionic gated \mote\ transistors, as expected (the value of $S$ is limited by disorder in the MoTe$_2$ crystals, and in cleaner materials --e.g., \ws-- the value measured is close to 60 mV/decade).}
	\label{fig:2}
\end{figure}

The use of electrolytes to transfer the potential from a gate electrode to a semiconductor surface started in the earliest days of modern electronics, with the development of the transistor (see the first of John Bardeen’s two Nobel lectures\cite{bardeen_john_semiconductor_1956} and Ref.~\cite{brattain_experiments_1955}). Electrostatically, an electrolyte behaves like a metal, so that the potential difference applied between two electrodes immersed in its interior is fully screened within a short distance\cite{kim_electrolyte-gated_2013,fujimoto_electric-double-layer_2013,bisri_endeavor_2017}, past which the electrolyte is equipotential (see Fig. \figref{fig:1}{b}). For the ionic concentrations of interest here, the \emph{Debye} screening length is 1~nm or smaller, making the electrode/electrolyte interfaces across which the voltage drops behave as capacitors with an enormous capacitance per unit area, $C_G \approx$~20-60~$\mu$F/cm$^2$.\cite{ono_comparative_2009,kim_electrolyte-gated_2013,fujimoto_electric-double-layer_2013,bisri_endeavor_2017} This is three orders of magnitude larger than the capacitance associated to a 300~nm thick layer of \sio\ commonly used in many experiments on two-dimensional materials (Fig.~\ref{fig:2})\cite{novoselov_two-dimensional_2005,geim_rise_2007}.\\

Different types of electrolytes have been employed for electrostatic gating. Work initially relied on water-based electrolytes\cite{brattain_experiments_1955,kruger_electrochemical_2001,rosenblatt_high_2002}, but it is now clear that if water --or even just traces of humidity-- is present in the electrolyte, the chemical stability of the material to be gated is easily affected. Progress in selecting suitable electrolytes and in refining different aspects of the technique came from research on organic  transistors\cite{nilsson_bi-stable_2002,panzer_low-voltage_2005,shimotani_direct_2005,misra_electric_2007,kim_electrolyte-gated_2013}. The use of ionic liquids (also in gel form\cite{kim_electrolyte-gated_2013}) as chemically stable electrolytes for gating\cite{nilsson_bi-stable_2002,panzer_low-voltage_2005,shimotani_direct_2005,misra_electric_2007}, or the implementation of a reference electrode\cite{xia_correlation_2009} to measure the electrostatic potential in the interior of the liquid, provide excellent examples. These and other advances brought ionic gating from being viewed as a tantalizing idea plagued by technical difficulties and artifacts, to an experimental technique with great potential.\\

During the last decade, ionic gating developed further through research on inorganic systems, with the main aim to control and investigate the electronic state of materials through charge accumulation\cite{ueno_electric-field-induced_2008,yamada_electrically_2011,ueno_discovery_2011,sohier_enhanced_2019}. Remarkable results in this context include the already mentioned demonstration of gate-induced superconductivity in semiconducting TMDs\cite{ye_liquid-gated_2010,ye_superconducting_2012,jo_electrostatically_2015,costanzo_gate-induced_2016,costanzo_tunnelling_2018,lu_full_2018,piatti_multi-valley_2018,kouno_superconductivity_2018,piatti_ambipolar_2019} or the observation of gate-induced ferromagnetism in thin platinum films\cite{liang_electronic_2014} (for an overview of activities we refer to comprehensive recently published review articles\cite{fujimoto_electric-double-layer_2013,bisri_endeavor_2017}). Broadening the classes of materials used in conjunction with ionic gating led to further improvements in understanding of the technique. It is now clear, for instance, that phenomena of non electrostatic origin can take place in ionic gated transistors, whose nature and relevance depends on the specific material considered. One example is provided by transition metal oxides, with oxygen atoms that are frequently driven out of the material by the very large electric field present in ionic gated transistors\cite{jeong_suppression_2013,fete_ionic_2016}. In some cases, the phenomenon can be exploited to deterministically control the concentration of oxygen vacancies that act as dopants, enabling the introduction of charge carriers in the system. The mechanism, however, entails chemical modifications of the material itself, and is therefore profoundly different from electrostatic doping (for completeness, some transition metal oxides are known, for which ionic gating does has a purely electrostatic effect\cite{scherwitzl_electric-field_2010,wang_scattering_2020}; for a good overview, see the introduction of Ref.\cite{wang_scattering_2020}). For other classes of compounds, ionic gating has a purely electrostatic effect as long as the applied gate voltage is kept within a suitable range. Among others, this is the case for virtually all semiconducting TMDs that are a main focus of this paper: for these materials ionic gating does not cause any chemical modification, and allows proper transistor operation with a high level of reproducibility. This is essential to implement ionic gate spectroscopy, which is what we discuss here. \\

\section*{Measuring the band gap of two-dimensional semiconductors}

Currently, the determination of the band gap of two-dimensional semiconductors mostly relies on either optical\cite{chernikov_exciton_2014,ruppert_optical_2014,klots_probing_2014,lezama_indirect--direct_2015,hanbicki_measurement_2015,poellmann_resonant_2015,wang_giant_2015,hill_observation_2015,raja_coulomb_2017,yore_large_2017} or scanning tunneling spectroscopy\cite{lu_bandgap_2014,zhang_direct_2014,chiu_determination_2015,huang_bandgap_2015,rigosi_electronic_2016,hill_band_2016}, two techniques that require modelling of the measured data to extract quantitative gap values. The main difficulty with optical spectroscopy experiments is that --irrespective of the precise quantity that is measured (reflectivity, absorption, photoluminescence)-- the signal is dominated by excitonic effects\cite{chernikov_exciton_2014,he_tightly_2014,ugeda_observation_2014,wang_colloquium_2018}. In contrast to the case of bulk semiconductors, the much stronger electron-hole interaction in atomically thin layers prevents the exciton recombination energy to be taken as a precise measure of the gap. In monolayers of semiconducting TMDs, the problem was solved through the analysis of excited exciton states that are observed in sensitive measurements on sufficiently high-quality devices. By assuming a realistic potential for the electron-hole interaction, the energy of these excited electron-hole states can be fitted and extrapolated to the continuum, whose onset corresponds to the energy of inter-band transitions between non-interacting particles\cite{chernikov_exciton_2014,he_tightly_2014} (\ie\ to the band gap of the material). For monolayers of the most common semiconducting TMDs, this approach results in very reasonable gap values. It is however not obvious whether the same approach can be extended to bilayers or thicker multilayers, which are indirect band gap semiconductors with weaker coupling of electronic transitions to light, and hence with a much weaker signal from exciton excited states\cite{splendiani_emerging_2010,mak_atomically_2010,zhao_evolution_2013}.\\

Even with scanning tunneling spectroscopy (STM), the determination of semiconducting band gaps is more complex than it may be anticipated. With gaps that are commonly larger than 1~eV, the bias that needs to be applied to the STM tip strongly modifies the probability of electron tunneling through vacuum, as well-known from bulk semiconductors\cite{stroscio_scanning_1993}. The resulting effect on the tunneling current is exponentially strong and needs to be modelled to extract reliable gap values from the measured differential conductance\cite{rigosi_electronic_2016}. Another complication originates from the electrostatic effect of the tip bias, which acts as a local gate that shifts the energy of the band edges\cite{feenstra_tunneling_1987,ugeda_observation_2014,hill_band_2016}. Finally, most work on atomically thin semiconductors is done on exfoliated material or nano-fabricated structures, whose surfaces are not prepared in ultra-high vacuum, significantly increasing the complexity of the experiments. For all these reasons, the determination of band gap of two-dimensional semiconductors by scanning tunneling spectroscopy is not straightforward and rather time consuming.\\

Other techniques can also be employed to determine the size of the band gap of two-dimensional materials, but the level of experimental complexity increases\cite{park_direct_2018}. Beautiful work, for instance, has been done by means of angle-resolved photoemission spectroscopy (ARPES) in two-dimensional semiconductors\cite{wilson_determination_2017,cucchi_microfocus_2019,hamer_indirect_2019} whose conduction band is populated by adding electrons through the deposition of alkali metal atoms on the surface\cite{riley_negative_2015,kim_possible_2017,kang_universal_2017,katoch_giant_2018} or --very recently-- by means of electrostatic gating\cite{nguyen_visualizing_2019}. This work is certainly of extreme interest to extract direct information about many aspects of the band structure of two-dimensional materials. The technical complexity and advanced instrumentation required to perform this type of measurements, however, are not comparable to techniques such as optical spectroscopy or scanning tunneling spectroscopy.

\section*{Spectroscopy of band edges in semiconductors: the idea}

The idea to exploit the very large geometrical capacitance of field-effect transistors to perform spectroscopy has repeatedly attracted attention in the past. The band gap of a semiconducting nanotube\cite{rosenblatt_high_2002}, for instance, was extracted from conductance measurements as a function of source-drain and gate bias, through a type of analysis normally done for Coulomb-blockaded quantum dots\cite{kouwenhoven_electron_1997}. Filling of a second sub-band was also detected\cite{ilani_measurement_2006,shimotani_continuous_2014}, and attempts were made to estimate its position in energy. Similarly, the onset voltage for hole transport in ionically gated organic polymer transistors was found to correlate with the energy position of the HOMO level of the specific polymer investigated, although no quantitative relation was established\cite{xia_correlation_2009}. Albeit interesting, these results were based on \emph{ad-hoc} assumptions for each system investigated, and no attempt was made to establish clearly which physical quantities can be reliably extracted quantitatively from the measurements, and under what conditions.\\

The situation has changed with the advent of two-dimensional materials, because the large number of semiconducting compounds available\cite{mounet_two-dimensional_2018} --all having different band gaps-- has allowed a spectroscopic strategy to be validated systematically. To understand the basic idea of spectroscopy based on ionic gated transistors\cite{braga_quantitative_2012}, recall that a shift in gate voltage  causes a shift in both the electrostatic potential ($\Delta \Phi$), and in the position of the chemical potential relative to the band structure ($\Delta\mu$), related as:
\begin{equation}
e\Delta V_G = \Delta\mu + e\Delta\Phi = \Delta\mu + \frac{e^2\Delta n}{C_G},
\label{eq:1}
\end{equation}
($e\Delta n$ is the charge density accumulated electrostatically due to the shift in gate voltage; see Fig.~\figref{fig:2}{a}). For a semiconductor with $\mu$ located inside the band gap, $\Delta n$ is small because electrons can be hosted only in defect states, whose concentration is low if the material quality is good --\ie\ the DOS inside the gap of a semiconductor is small. The last term in Eq. \ref{eq:1} is then negligible because $\Delta n$ is small and $C_G$ extremely large, and we obtain $e\Delta V_G \simeq \Delta\mu$, a relation valid for an ionic gated field-effect transistor with extremely good approximation as long as the chemical potential is located inside the gap (Fig.~\figref{fig:2}{a}, bottom panel).\\

\begin{figure}
	\centering
	\includegraphics[width=.79\textwidth]{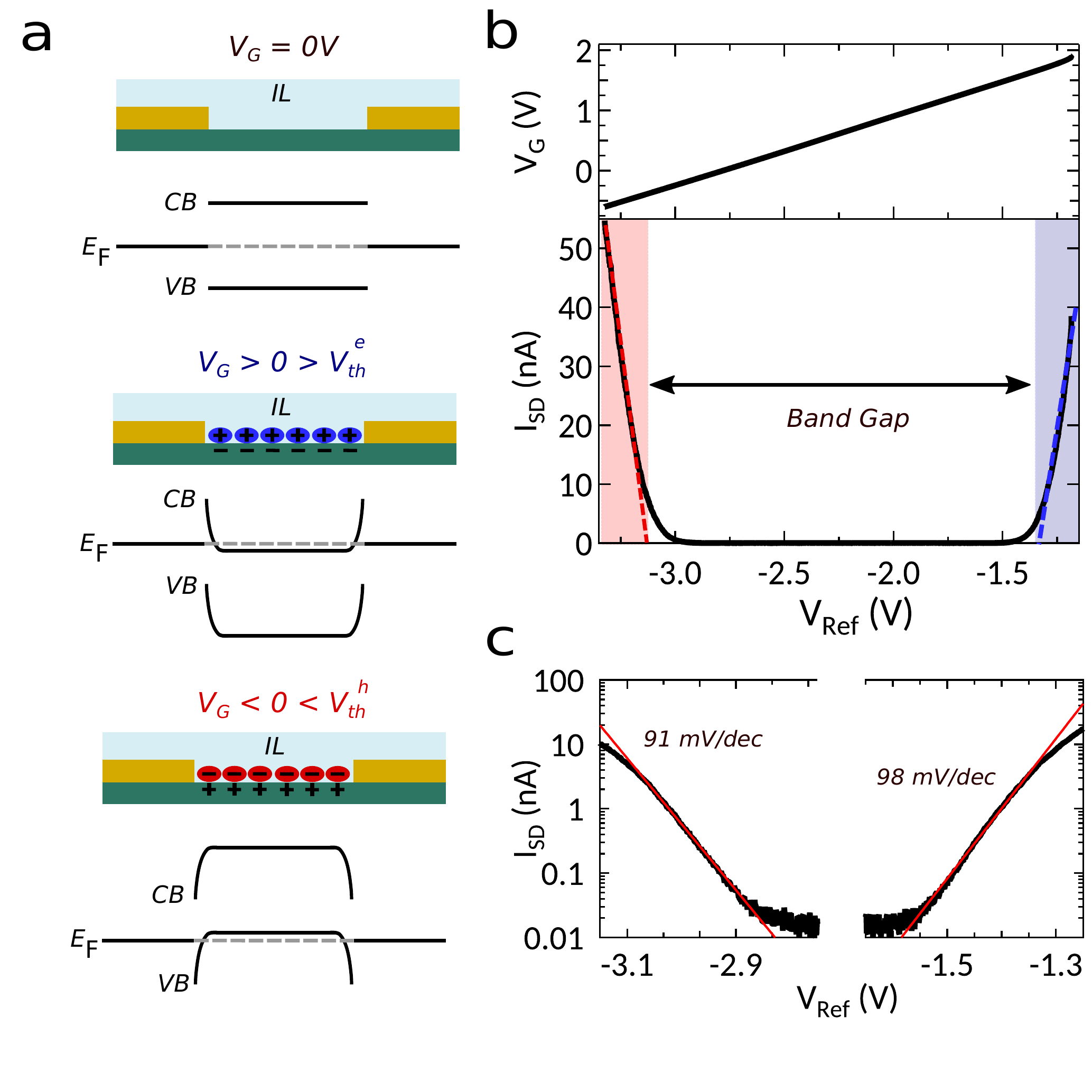}
	\caption{{\bfseries Determination of the band gap.} \footnotesize {\bfseries a}. Band diagram of a semiconducting channel in a field-effect transistor, for different values of gate voltage. When the transistor is in the off-state, the chemical potential is in the gap of the semiconductor (top panel). A positive gate voltage pushes the bands in the semiconductor to lower energy. Electrons from the contacts enter the transistor channel when the conduction band edge is aligned with the Fermi level in the contacts at $V_G$~=~\vthe\ (middle panel). A negative gate voltage pushes the bands in the semiconductor to higher energy and hole accumulation starts at $V_G$~=~\vthh, when the valence band edge aligns to the Fermi level in the contacts (bottom panel). {\bfseries b}. Top panel: linear relation between the applied gate voltage $V_G$ and the reference potential $V_{Ref}$, with an offset  due to the built-in potential. Bottom panel: Ambipolar transfer curve \isd($V_G$) of a monolayer \wse\ transistor. The light red and blue shaded regions correspond to the regimes in which transport is mediated respectively by holes in the valence band or by electrons in the conduction band. The dashed lines  illustrate how the threshold voltages \vthe\ and \vthh\ are determined, from which the  gap $\Delta = e(\vthe - \vthh))$~=~1.8~eV is extracted (see also Box 2). {\bfseries c}. The small value of  subthreshold slope $S$ for both hole and electron accumulation indicates an excellent electrostatic coupling between gate and transistor channel.}
	\label{fig:3}
\end{figure}

For transistors built on atomically thin two-dimensional semiconductors, the charge accumulated electrostatically in the channel is transferred from the source/drain metal electrodes that are virtually equipotential (if the source-drain bias \vsd\ is small, as it is the case for the discussion here; Fig.~\figref{fig:3}{a}). Transfer of electrons into the channel starts at $V_G = \vthe$, the threshold voltage for electron conduction, when the applied gate voltage causes the conduction band edge in the semiconductor to align to the chemical potential in the contacts (Fig.~\figref{fig:3}{a}, mid panel)\cite{podzorov_high-mobility_2004}. Similarly, hole accumulation first occurs at $V_G = \vthh$ (\ie\ the threshold voltage for hole conduction) when the gate voltage lifts the valence band edge to align to the chemical potential in the contacts (Fig.~\figref{fig:3}{a}, bottom panel). As $V_G$ is shifted from \vthh\ to \vthe, the condition $e\Delta V_G = \Delta\mu$ holds true at all times, because the chemical potential in the semiconductor always remains inside the gap. It then directly follows that $e(\vthe - \vthh)$ corresponds quantitatively to the difference in energy between the conduction and valence band edge, which is --by definition-- the band gap $\Delta$ of the semiconductor (Fig.~\figref{fig:3}{b}). It is this relation that provides a straightforward strategy to extract the band gap of a semiconductor from the ambipolar transport characteristics of an ionic gate transistor.\\

A comment is worth as to whether the electric field perpendicular to the semiconductor --unavoidably present in a field-effect transistor under charge accumulation-- may cause problems to the validity of our argument. Specifically, a large perpendicular electric field can strongly deform the bands in the semiconductor and change the gap, as it is well established for the case of TMD multilayers\cite{ramasubramaniam_tunable_2011,brumme_first-principles_2015}. One  may therefore wonder whether the value of gap that is extracted by ionic gating spectroscopy does correspond to that  of the pristine material, or whether it is influenced by the experimental conditions. The key issue to answer this question is that the gap is determined as the difference of threshold voltage for electron and hole conduction. For $V_G$ close to $V_{\rm th}$, the density of charge accumulated on the semiconductor is small (it vanishes at threshold for the ideal case in which no defect states are present in the gap) and so is the electric field perpendicular to the transistor channel. It is for this reason that no significant band modification is taking place under the condition of the experiment, and that the extracted gap value does indeed correspond to the pristine gap of the material investigated.\\

On the experimental side, our discussion  assumes the voltage applied to the gate to drop entirely at the interface between the electrolyte and the semiconductor. If the gate leakage current is negligible, this is the case if the capacitance between gate and electrolyte is much larger than that between electrolyte and semiconductor, which is why in actual devices the gate electrode is designed to have a very large area. If the voltage drop at the gate/electrolyte interface is not negligible (see Fig.~\figref{fig:1}{b}), the discussion above remains correct provided that we substitute $V_G$ with \vref, the voltage measured at the reference electrode (Fig.~\figref{fig:3}{b}). That is because the reference electrode --being in equilibrium with the interior of the electrolyte-- measures directly the voltage across the electrolyte/semiconductor interface, irrespective of any voltage drop at the gate/electrolyte interface\cite{shimotani_gate_2006,xia_correlation_2009}. The use of a reference electrode, therefore, eliminates effects due to spurious phenomena occurring at the gate/electrolyte interface and allows more precise quantitative estimates of the gap. For instance, the hysteresis in the source-drain current observed in certain devices when plotting \isd\ versus $V_G$ disappears when \isd\ is plotted as a function of \vref\cite{gutierrez-lezama_electroluminescence_2016}. In practice, checking that a change $\Delta V_G$ in the potential applied to the gate induces a change $\Delta V_{Ref}$ of very close magnitude is also a first, simple and effective way to monitor that the device operates correctly. Despite the practical relevance of using a reference electrode, to streamline the discussion, we will assume hereafter that the voltage drop at the gate/electrolyte interface is negligible, and treat $V_G$ and \vref\ as equivalent unless the opposite is explicitly mentioned.\\

It should be noted that the band gap is extracted from the difference of two values of $V_G$ measured on a same device (see Fig.~\figref{fig:3}{b}), and one may wonder whether individual values of $V_G$ (\ie\, not only their difference) also have a well-defined physical meaning. At least in principle this is the case, because threshold voltages for electron and hole conduction are determined by physical quantities --such as the work functions of the source/drain contacts, of the gate electrode, or the doping concentration of the semiconductor forming the channel-- that are supposedly the same in nominally identical devices. One could then envision compiling a catalogue of threshold voltages for different semiconductors, extracted from measurements on different devices equipped with a same \emph{conventional} reference electrode, in the same spirit of how electron-transfer reactions are catalogued in electrochemistry\cite{bard_electrochemical_nodate}. In practice, however, the level of experimental control is not sufficient to make a meaningful comparison, because threshold voltages measured on different devices exhibit offsets of extrinsic nature. Their most common origin is variations in the materials work function, which is strongly affected by the condition of the material surface, and can cause sizable device-to-device fluctuations in the value of the built-in electrostatic potential difference between the gate and the transistor channel. Comparing energies in a same device --which is what the relation $\Delta = e(\vthe - \vthh)$ does-- eliminates the issue, because the offset in electrostatic potential cancels out when taking the difference of two threshold voltages. Reliable and reproducible result, therefore, can only be obtained for physical quantities that can be extracted from the difference of gate voltages measured on a same device.\\ 

\FloatBarrier
\section*{Determination of band gaps}

\begin{figure}[ht]
	\centering
	\includegraphics[width=.95\textwidth]{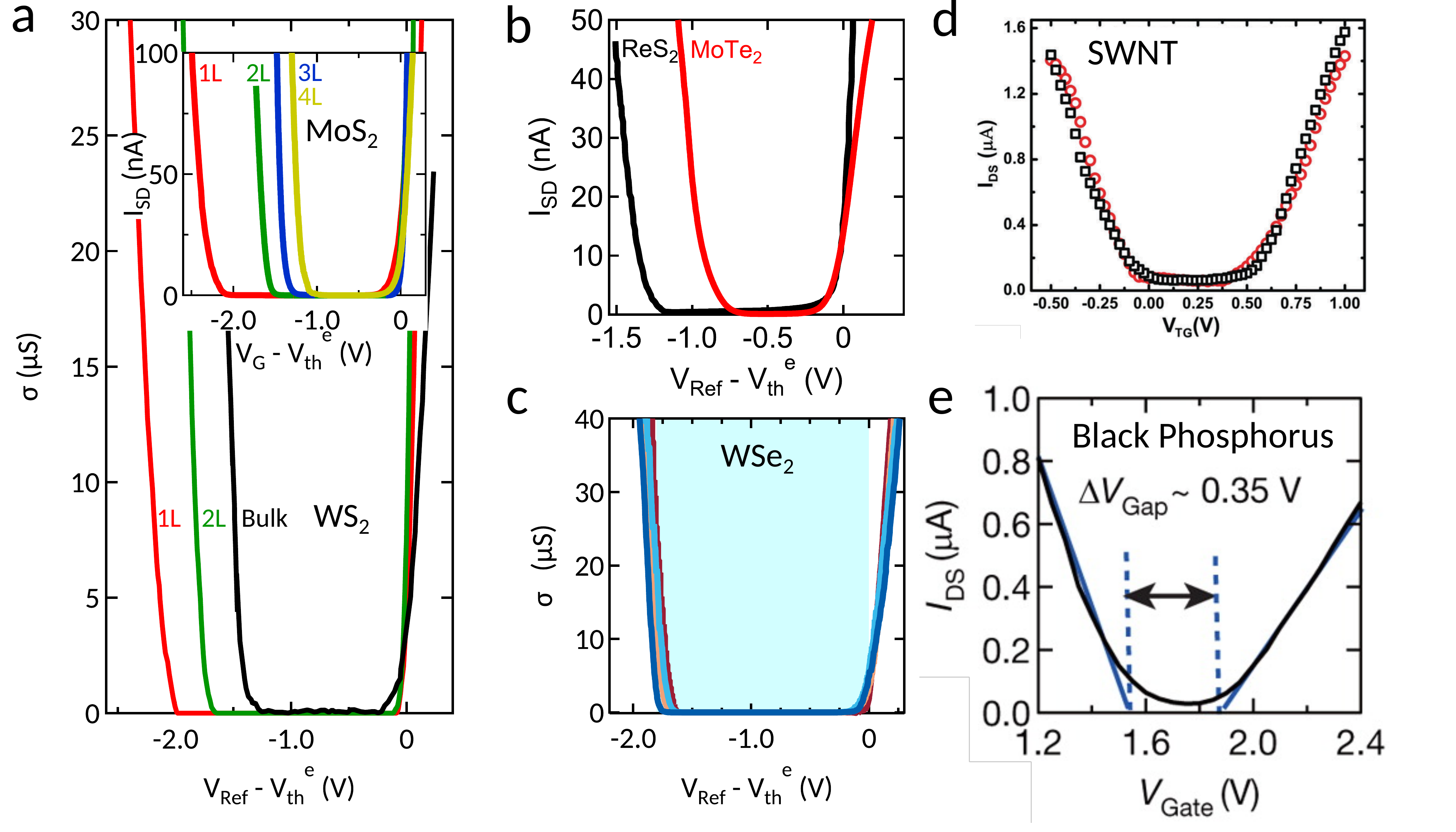}
	\caption{{\bfseries Precision and reliability of ionic gate spectroscopy.} \footnotesize {\bfseries a}. Ambipolar transfer curves of ionic gated transistors based on monolayer (red curve), bilayer (green curve), and bulk (a 10 nm exfoliated layer; black curve) \ws\ crystals, giving the following values for the corresponding band gaps:  $\Delta_{1L}$~=~2.12~eV, $\Delta_{2L}$~=~1.82~eV, and $\Delta_{Bulk}$~=~1.4~eV, in agreement with expectations (adapted from Refs~\cite{braga_quantitative_2012} and \cite{jo_mono-_2014}). The inset shows similar curves for mono-, bi-, tri-, and tetralayer \mos\, from which $\Delta_{1L}$~=~2.55~eV, $\Delta_{2L}$~=~1.6~eV, $\Delta_{3L}$~=~1.4~eV, and $\Delta_{4L}$~=~1.25~eV (adapted from Refs.~\cite{ponomarev_ambipolar_2015} and \cite{ponomarev_hole_2018}). All data are measured on exfoliated layers, except for the monolayer, in which experiments have been done on a transistor realized on a CVD-grown material (see discussion in the main text). {\bfseries b.} From ambipolar transfer curves, the gap of thick exfoliated MoTe$_2$ (red curve) and ReS$_2$ (black curve)  crystals is found to be $\Delta$\mote\~=~0.85~eV and $\Delta$ReS$_2$~=~1.4~eV, as expected  (adapted from Refs.~\cite{gutierrez-lezama_electroluminescence_2016} and \cite{lezama_surface_2014}). {\bfseries c} Transfer curves of five different monolayer \wse\ transistors realized using two different ionic liquids. The gap value extracted from the  different devices, ranging from 1.75~eV to 1.87~eV, illustrates the precision and reliability of the technique ($< \pm 5$\% uncertainty). {\bfseries d.} and {\bfseries e.} Ionic liquid gated transistors based on respectively a semiconducting single wall nanotube and on a thick exfoliated black phosphorous layer (adapted from Refs.~\cite{saito_ambipolar_2015} and \cite{vasu_opening_2018}).}
	\label{fig:4}
\end{figure}

The possibility to extract precise band gap values from ionic gated transistor measurements first came as a surprise during work on ambipolar transport in \ws\ devices\cite{braga_quantitative_2012}. In these devices --based on rather thick exfoliated layers, behaving electronically as bulk-- the difference between electron and hole threshold voltage was systematically found to be close to 1.4~eV, matching quantitatively the value of the band gap of bulk \ws\ (see Fig.~\figref{fig:4}{a}). These experiments were correctly explained in terms of an ideal electrostatic coupling between gate and transistor channel, but it was initially believed that such an excellent agreement was exclusively due to the exceptional quality of the \ws\ crystals used in those experiments (grown by Helmuth Berger at EPFL by an extremely slow vapor phase transport process). At that stage, it was considered unlikely that other materials with a larger concentration of defects would allow an equally precise determination of the gap.\\

The results obtained on \emph{bulk-like} \ws\ were confirmed by similar measurements performed later on mono- and bilayers exfoliated from the same crystals\cite{jo_mono-_2014}. Equally good ambipolar transport was observed on these atomically thin crystals, from which gap values fully consistent with expectations were extracted, respectively 2.12~eV and 1.82~eV for monolayer and bilayer \ws\ (see Fig.~\figref{fig:4}{a}). It was not until experiments were done on ReS$_2$\cite{gutierrez-lezama_electroluminescence_2016} and \mote\cite{lezama_surface_2014} field-effect transistors, however, that the robustness of ionic gating spectroscopy became fully apparent. Crystals of both these materials had not been grown following any special procedure and different experimental observations --hysteresis in the transfer curve due to bias-stress effects, deviations from the ideal value of subthreshold slope, and direct spectroscopic imaging by means of scanning tunneling microscopy-- indicated the presence of a considerable density of in-gap states. Nevertheless, ionic gate transistor measurements showed reproducible results (Fig.~\figref{fig:4}{b}), leading to band gap values in excellent agreement with estimates obtained by either STM or optical spectroscopy measurements (which for bulk or thick exfoliated crystals is a good term of comparison). Comparable gap values for \mote\ were also reported later by a different group\cite{xu_reconfigurable_2015}, using ionic gated transistors in conjunction with a procedure slightly different from the one discussed here, confirming that the results do not depend on specific details of the ionic gated transistors employed or of their operation.\\

The robustness and accuracy of ionic gating as a spectroscopic technique have been subsequently confirmed by numerous other experiments\cite{saito_ambipolar_2015,ovchinnikov_disorder_2016,prakash_bandgap_2017,ciarrocchi_thickness-modulated_2018}. One is the measurement of the systematic evolution of the band gap with increasing number of layers, from mono (1L) to four-layer (4L), in \mos\cite{ponomarev_ambipolar_2015,ponomarev_hole_2018}. For \mos, differently from other common TMDs, exfoliated monolayers do not exhibit ambipolar transport because of a high density of defect states in the gap (caused by sulfur vacancies) that act as traps for holes\cite{ponomarev_hole_2018}. It is nevertheless possible to observe ambipolar transport (see inset of Fig.~\figref{fig:4}{a}) and extract the band gap by measuring monolayer devices realized on material synthesized by chemical vapor deposition (CVD) that --being grown in a strong excess of sulfur-- contains less vacancies. The gap was found to be 2.55~eV for 1L-\mos\ \cite{ponomarev_ambipolar_2015}(see discussion below as to why the value is somewhat lager than the expected one), 1.6~eV for 2L-\mos, 1.4~eV for 3L-\mos, and 1.25~eV for 4L-\mos\cite{ponomarev_hole_2018}. Fig.~\figref{fig:4}{a} shows that the difference in band gap between 3L and 4L \mos\ is very easily appreciable in the experiments, despite the difference being less than 200 meV. The technique, therefore, has sufficient sensitivity to discriminate between gap values that differ by only a few tens of meV, over an absolute magnitude of approximately 1.5~eV, corresponding to an uncertainty of approximately 5\%.\\

\begin{figure}[ht]
	\centering
	\includegraphics[width=.95\textwidth]{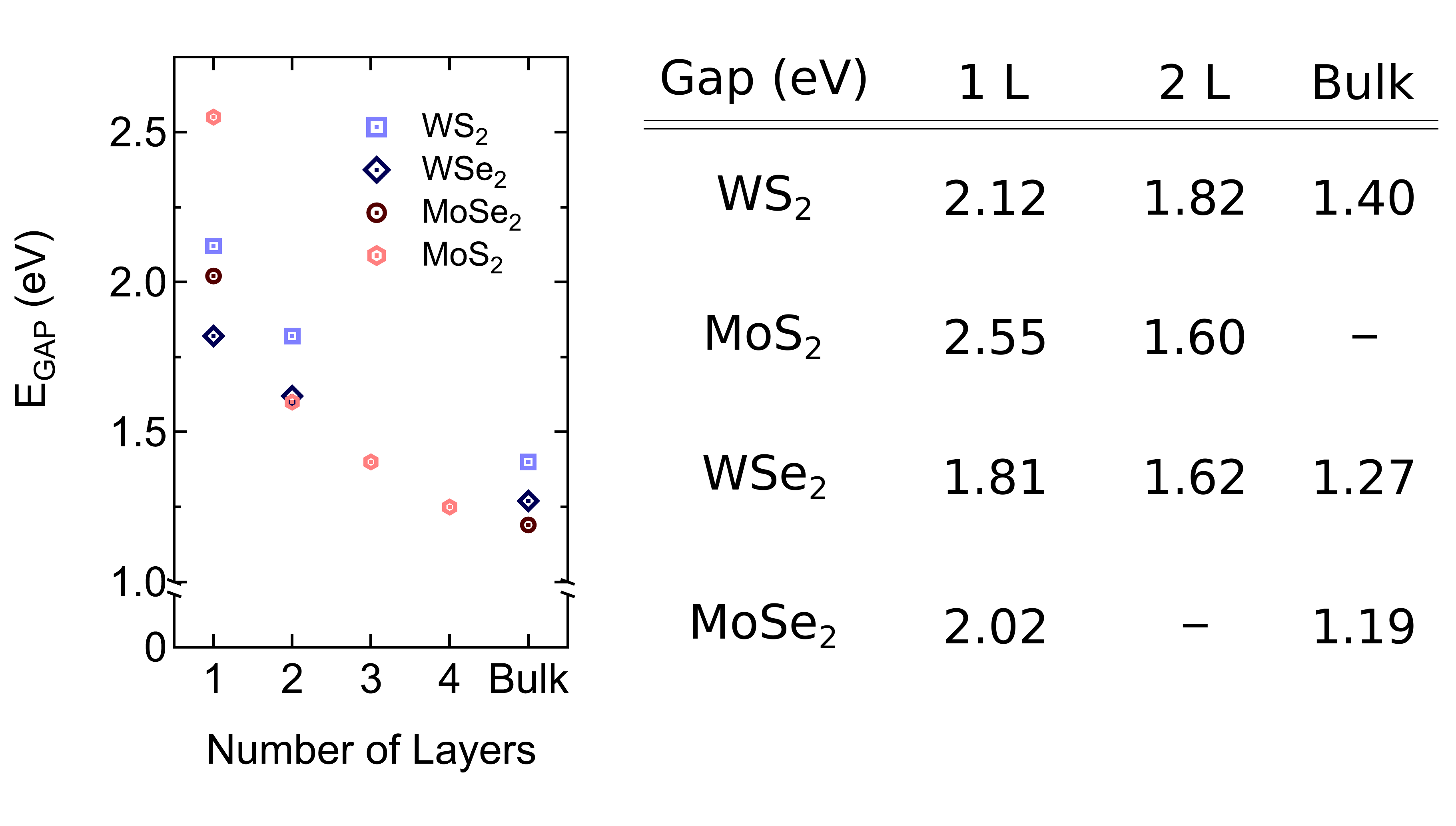}
	\caption{{\bfseries Gap values for common semiconducting TMDs.} \footnotesize Magnitude of the band gap of the most commonly used semiconducting TMDs (\ws, blue squares; \wse, black diamonds; \mose, brown circles; \mos, red hexagons) for different thicknesses, as determined by ionic gate spectroscopy. The table summarizes the values of the band gap measured in our group for monolayer, bilayer and bulk crystals of these compounds. All measurements have been performed on exfoliated layers, except for monolayer \mos, for which CVD-grown material has been used. The resulting value of the gap for monolayer MoS$_2$ is somewhat larger than expected due to a large density of trap states in the gap, at approximately 300~meV above the valence band edge (see discussion in the main text).}
	\label{fig:5}
\end{figure}

To establish the reproducibility of the gaps values extracted by ionic gate spectroscopy, we repeated measurements of nominally identical devices made with different two-dimensional semiconductors. Fig.~\figref{fig:5}{c} illustrates these results with ambipolar transfer curves measured on different \wse\ monolayer devices\cite{zhang_band_2019}. The curves exhibit only very small device-to-device fluctuations, despite the fact that two different electrolytes (DEME-TFSI and P14-FAP) were employed (several more devices were realized, all giving results consistent with those illustrated in the figure). By applying the relation $\Delta = e(\vthe - \vthh)$ to all of the curves we find values of the gap ranging from 1.75~eV to 1.87~eV, \ie\ $\Delta = 1.81 \pm 0.06$~eV, corresponding to an accuracy better than  $\pm 5$\% . Such a small uncertainty indicates that ionic liquid gating spectroscopy performs extremely well as a technique to determine the absolute value of the band gap in semiconductors. The gap of materials other than TMDs has also been obtained with this technique, as illustrated in Fig.~\figref{fig:5}{d} and \figref{fig:5}{e} with measurements on individual carbon nanotubes\cite{vasu_opening_2018} and on black phosphorous\cite{saito_ambipolar_2015}. Another example is discussed in Ref.~\cite{yomogida_ambipolar_2012} in which ionic gating is applied to organic crystals of rubrene and pentacene. In that case, however, the need to use large and different source drain-bias voltage to detect electron and hole transport (due to contact resistance problems that can be rather severe in organic semiconductors) casts doubts as to the quantitative validity of the estimation of the band gap from the difference of the respective threshold voltages (because the threshold voltage depends on the source-drain bias).\\

A systematic overview of gap values extracted from ionic liquid gated devices based on monolayers, bilayer, and bulk crystals of the four most commonly used semiconducting transition metal dichalcogenides (\mos, \mose, \ws, and \wse) is shown in Fig.~\ref{fig:4}. The size of the gap systematically decreases with increasing thickness for all compounds, and values are consistent with other measurements. For instance, whenever an excitonic transition is observed in optical spectroscopy measurements (e.g., photoluminescence), the energy of the exciton is smaller than the gap extracted from ionic liquid gating \cite{jo_mono-_2014}. The gap values are also consistent with values reported in the literature, but this agreement does not always pose a stringent constraint, because in a number of cases the values reported in the literature span a rather broad range of energies\cite{chernikov_exciton_2014,ruppert_optical_2014,klots_probing_2014,hanbicki_measurement_2015,poellmann_resonant_2015,wang_giant_2015,hill_observation_2015,raja_coulomb_2017,yore_large_2017,lu_bandgap_2014,zhang_direct_2014,chiu_determination_2015,huang_bandgap_2015,rigosi_electronic_2016,hill_band_2016} (much broader than the range of values obtained from measurements done on distinct ionic gated devices). In performing quantitative comparisons one should recall that the gap --as well as other quantities that can be extracted from it-- can depend on the dielectric environment because of screening effects that result in relatively large quantitative changes (\eg\ the gap of monolayer TMDs has been reported to differ by several hundreds of meV in systems exposed to different dielectric screening conditions)\cite{rosner_two-dimensional_2016,raja_coulomb_2017,cho_environmentally_2018,waldecker_rigid_2019}.\\

\FloatBarrier
\section*{Band alignment in different two-dimensional materials}

\begin{figure}[ht]
	\centering
	\includegraphics[width=.45\textwidth]{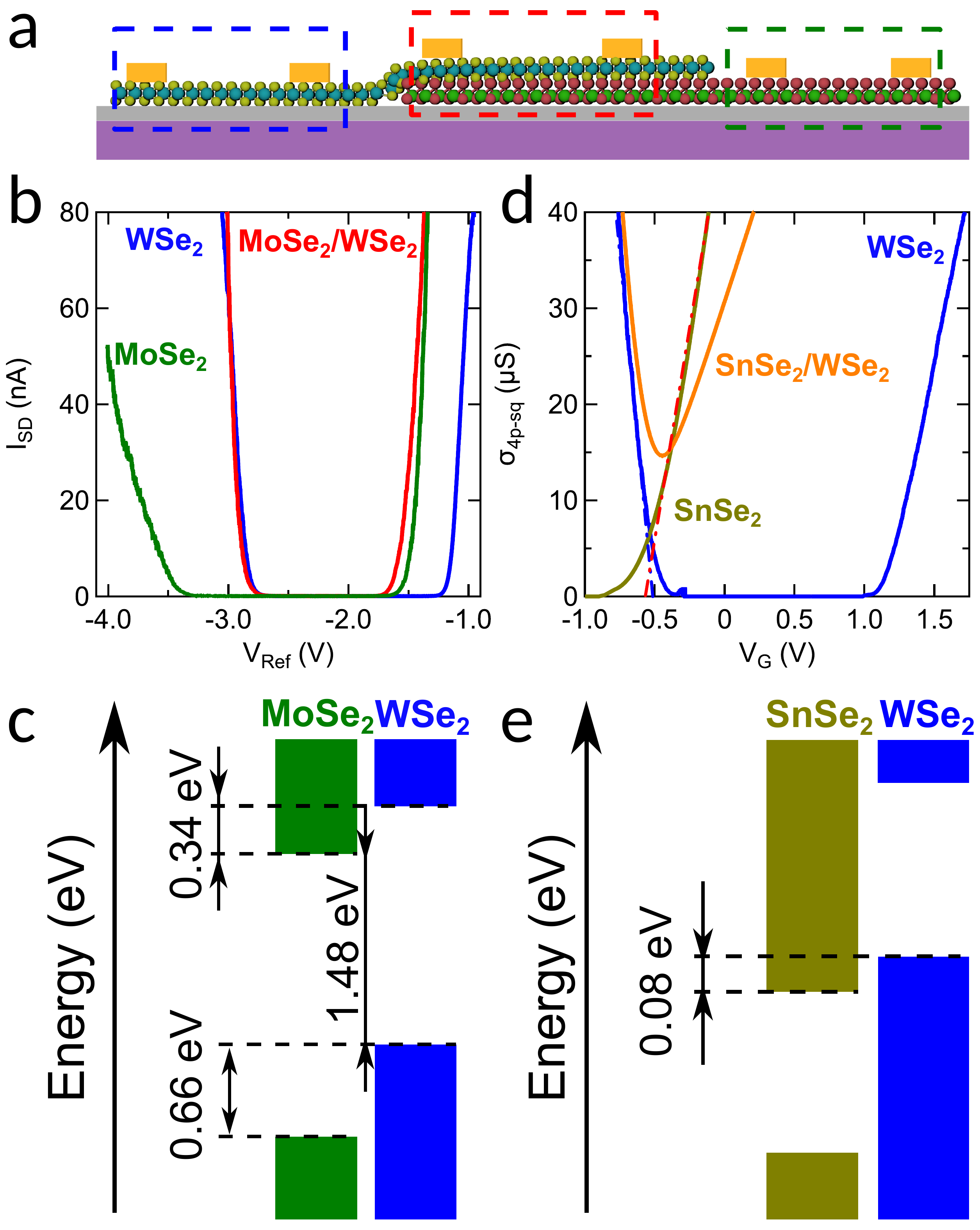}
	\caption{{\bfseries Band alignment in different semiconductors. } \footnotesize {\bfseries a.} The determination of the relative position in energy of the band edged of two different 2D semiconductors --i.e., their band alignment-- relies on structures that establish electrical contact between the two materials, while allowing independent transistor measurements to be performed on each one of them, as shown in the scheme. {\bfseries b} Transfer curves measured on the three different parts of a structure such as the one shown in {\bfseries a}, in which the two semiconductors are \wse\ and \mose\ monolayers (adapted from Ref.~\cite{ponomarev_semiconducting_2018}). The blue (green) curve are measurements done on the \wse\ (\mose) parts of the devices. The differences of the electron(hole) threshold voltages gives the offset between the conduction(valence) band edges in the two materials, equal to 0.34~eV (0.66~eV). The transfer curve of the interface (red line) overlaps with the curve measured on \mose\ for electron transport and with that  measured on \wse\ for hole transport, as expected for type-II alignment. {\bfseries c.} Diagram summarizing the band alignment between \mose\ and \wse\ monolayers. {\bfseries d.} Same measurements as in panel {\bfseries b} on a structure of the type shown in {\bfseries a}, based on three-layer \wse\ and four-layer SnSe$_2$ (adapted from Ref.~\cite{reddy_synthetic_2020}). The olive (blue) line represents the transfer curve measured on the SnSe$_2$ (\wse) part. The electron threshold voltage in SnSe$_2$ is 80 meV smaller than the hole threshold voltage in \wse, directly showing that the valence band edge in \wse\ is at a higher energy than the conduction band edge in SnSe$_2$. In the interfacial region (orange curve)  the source-drain current remains finite throughout the entire $V_G$ interval. The interface conductivity coincides with that of WSe$_2$ for sufficiently large negative $V_G$, but it is smaller than the conductivity of SnSe$_2$ for positive $V_G$. That is because in the interface region the SnSe$_2$ layer is separated from the ionic liquid by a WSe$_2$ trilayer, which is is thicker --and has a significantly smaller geometrical capacitance -- than the hBN monolayer that separates SnSe$_2$ from the ionic liquid in the region away from the interface (see Ref.~\cite{reddy_synthetic_2020} for details). {\bfseries e.} Band diagram of the trilayer \wse/tetralayer SnSe$_2$ system.}
	\label{fig:6}
\end{figure}

The same strategy followed to determine the band gap of two-dimensional semiconductors can be employed to extract the relative energetic position of the band edges in different materials, \ie\ their so-called band alignment. The measurements rely on devices fabricated on \emph{van der Waals} heterostructures\cite{geim_van_2013,novoselov_2d_2016} in which the two materials are electrically connected together, while still allowing gate-dependent transport to be measured on each of them independently (contacts deposited on top of the two individual materials, as well as on top of their interface, enable the current through each part of the structure to be measured separately; see Fig.~\figref{fig:6}{a}). As discussed earlier, because the two 2D materials are in direct contact, any extrinsic offset in electrostatic potential is the same for both of them and does not affect the measured difference of  threshold voltages. We have applied this strategy to show that the relative position of the band edges can be determined both for type-II and type-III heterostructures\cite{ponomarev_semiconducting_2018,reddy_synthetic_2020}.

Type-II band alignment occurs, for instance, in interfaces between \wse\ and \mose\ monolayers\cite{rivera_observation_2015,wilson_determination_2017}. Fig.~\figref{fig:6}{b} shows the source drain current (\isd) as a function of reference potential \vref\ measured on the \wse\ monolayer (blue line), the \mose\ part (green line), and on their interface (red continuous line).  For each of the monolayers --as well as for their interface-- the difference in threshold voltage for electron and hole conduction gives the band gap. The resulting value of the interfacial gap --1.48~eV-- is larger than the energy of the interlayer interfacial exciton --1.36~eV-- extracted from photoluminescence spectroscopy, leading to an exciton binding energy of 120 meV\cite{ponomarev_semiconducting_2018} (the expected order of magnitude\cite{wilson_determination_2017}). The \isd\ (\vref) curve measured on the interface coincides with that measured on the \mose\ monolayer for electron conduction and with that measured on the \wse\ monolayer for hole conduction (Fig.~\figref{fig:6}{b}), showing directly from the experiments that the conduction and valence band of the interface coincide with the corresponding bands of \mose\ and of \wse, as expected for a type-II heterostructure. Since all data are plotted as a function of \vref\ measured on the same reference electrode, the threshold voltages extracted from different parts of the structure can be quantitatively compared. That is how we extract the energetic distance between the band edges in the two different monolayers. As summarized in Fig.~\figref{fig:6}{c}, for  monolayer \mose\ and \wse\  we find  340~meV and 660~meV respectively for the conduction and valence band  edges. It is really quite remarkable that precise values of band offsets can be extracted straightforwardly by simply looking at the transfer curve of a transistor device.\\

Type-III band alignment occurs more rarely, and in vdW heterostructures of TMDs it was only recently observed in interfaces of few-layer \wse\ and SnSe$_2$\cite{aretouli_epitaxial_2016,yan_tunable_2017}. The structure employed to measure the overlap between the valence band of \wse\ and the conduction band of SnSe$_2$ is analogous to the one realized to determine the alignment of the bands in \mose\ and \wse. As the chemical stability of SnSe$_2$ is good but not perfect, the interface is realized placing the SnSe$_2$ multilayer under the \wse\ (the part of the SnSe$_2$ not covered by \wse\ is intentionally protected with a monolayer hBN, with the assembly of the \emph{van der Waals} heterostructure done entirely in the controlled atmosphere of a glove box). From measurements identical to those done for \mose\ and \wse\ interface, the threshold for electron conduction in SnSe$_2$ is found to occur at a smaller voltage than the threshold for hole conduction in \wse\ (see Fig.~\figref{fig:6}{d}), which signals that the edge of the SnSe$_2$ conduction band is lower in energy than the valence band edge in \wse. Consistently with this conclusion, different transport measurements performed on interfaces of SnSe$_2$ and \wse\ multilayers show the coexistence of electrons and holes\cite{reddy_synthetic_2020}. The magnitude of the overlap obtained from the measured threshold voltages is  80~meV, with a significant uncertainty originating from the fact that the band overlap is a small quantity obtained from the difference of two much larger ones (Fig.~\figref{fig:6}{e}). Note that the relation $\Delta \mu = e \Delta V_G $ does not hold true in the \wse/SnSe$_2$ interface itself, which is not gapped. However, it does hold for the constituent \wse\ and of SnSe$_2$ multilayers when the chemical potential is in the gap, and that is all what we need to extract the value of band overlap. Indeed, inasmuch measurements of the band alignment are concerned, the \wse/SnSe$_2$ interface region has the only role to keep the \wse\ and the SnSe$_2$ layer connected, to eliminate any relative extrinsic offset in the electrostatic potential of the two materials.\\

\FloatBarrier
\section*{Assessment of the technique}

The examples presented above illustrate how ionic gated transistors allow spectroscopic measurements of the relative position of band edges in semiconductors in a rather straightforward manner, and show the precision and reproducibility of the technique. Possibly the most important aspect of the technique is that the gap of two-dimensional semiconductors is extracted quantitatively through a procedure that is entirely well-defined: the gap value is obtained directly from the measurement of the threshold voltages for electron and hole conduction, without the need to model or fit the data. In contrast to what we discussed earlier for optical or scanning tunneling spectroscopy, therefore, ionic gate spectroscopy is a technique that directly measures the band gap. An incorrect value of the gap can result from either an incorrect experimental determination of the threshold voltage (see Box 2), from the presence of a non-negligible gate leakage current (which causes the ionic liquid not to be equipotential and prevents the analysis of the electrostatic potential in the transistor in terms of capacitances), or from the fact that the equality $e \Delta V_G = \Delta \mu$ is not satisfied as the chemical potential is shifted through the gap while sweeping the gate voltage. The latter situation is the reason why the gap value extracted for CVD \mos\ monolayer (see Fig.~\figref{fig:4}{}) is somewhat larger than the values obtained by other techniques\cite{ponomarev_ambipolar_2015}. Specifically, that is because close to the top of the valence band a rather high density of states inside the gap is present that can host a relatively large amount of charge\cite{ponomarev_hole_2018} making the shift in electrostatic potential not entirely negligible, \ie\ close to the top of the valence band the relation $e \Delta V_G = \Delta \mu$ does not exactly hold true in CVD \mos\ monolayers (these in-gap states originate from sulfur vacancies. They are specific to MoS$_2$ monolayers, because in monolayers of other common semiconducting TMDS --or in thicker layers of MoS$_2$-- the energy of the equivalent states originating from chalcogen vacancies falls inside the valence band and not in the band gap; see Ref.~\cite{ponomarev_hole_2018} for a more detailed discussion).\\

Another advantage of the technique is the possibility to make diagnostic tests on the same devices used to perform spectroscopic measurements. The key condition to perform spectroscopy with ionic gated devices is that the electrostatic coupling is ideal,\ie\ that a change in gate voltage induces an equal change in electrostatic potential of the semiconducting channel relative to the grounded electrodes. This condition can be tested (see Figs.~\figref{fig:1}{c} and \figref{fig:2}{c}), as it is related to the value of subthreshold slope $S$ in the device, a readily measurable quantity. For perfect coupling ($S = \frac{k_B T}{e} {\rm ln}(10)$, 60~mV/dec at room temperature), measurements on actual devices regularly give values close to this ideal limit. Our experiments show that even deviations from this limit of a factor of 2-3 do not prevent a precise determination of the band gap.\\

As for the limitations of the technique, the most serious one is that --as we mentioned-- ionic gate spectroscopy cannot be used on all materials, because not all materials are sufficiently stable chemically to be employed as channels of ionic gated transistors. The technique also cannot be applied to materials that do not exhibit gate-induced ambipolar transport, which may happen for different reasons. One is if the bands are extremely flat and cause electronic states to localize, as it is the case in some two-dimensional magnetic semiconductors of current interest\cite{gibertini_magnetic_2019} (\ie\ CrI$_3$ or CrCl$_3$). Another is the presence of a large density of in-gap states that host the electrostatically accumulated charge and pin the chemical potential, preventing it from being shifted into one of the bands. This is the case for exfoliated \mos\ monolayers\cite{ponomarev_hole_2018} and for InSe multilayers\cite{bandurin_high_2017,guo_band_2017}, in which only unipolar (electron) transport has been observed so far. Finally, it should be realized that the technique does not work at temperatures much lower than room temperature, since the electrolyte freezes, and the electrostatic potential does not respond to a change in the applied gate voltage. This does not pose problems for most materials of interest, whose gap is much larger than $k_B T$ at room temperature. For systems having a much smaller band gap, however, the limited energy resolution caused by the need to work close to room temperature may be more problematic.\\

\FloatBarrier
\section*{Conclusions and perspectives}

Ionic gate spectroscopy is a very effective technique to measure band gaps of semiconductors, as well as the relative alignment of bands in different semiconducting materials, in a precise and reliable way. The conditions that need to be fulfilled for the technique to work properly --most notably that the density of impurity state inside the band gap is small-- make layered \emph{van der Waals} bonded semiconductors the ideal systems for the application of ionic gate spectroscopy. That is because in layered \emph{van der Waals} materials the absence of broken covalent bonds at the surface normally eliminates in-gap states that very commonly occur at surfaces of conventional covalently bonded semiconductors, such as III-V compounds. The technique is therefore of great interest in the development of artificial structures with engineered electronic properties, such as those created by stacking on top of each other atomically thin \emph{van der Waals} crystals of different compounds\cite{geim_van_2013,novoselov_2d_2016}. Ionic liquid gating spectroscopy has an enormous potential as a tool to characterize the energetic position of the bands in the constituent materials, enabling some measurements that are not possible in an equally simple way with other techniques (\eg, the characterization of bi- and multilayers). It also offers the possibility to characterize the properties of the assembled structures, as already demonstrated in the simplest case of interfaces of two different atomically thin crystals.\\

Future research should aim at applying the technique to a broader class of systems and at exploring if and how ionic liquid gating can be used to obtain spectroscopic information also in the regime in which the chemical potential is shifted inside the bands of the system investigated (\ie\ not only across its band gap). First measurements performed on graphene mono-, bi- and trilayers\cite{ye_accessing_2011,gonnelli_weak_2017}, on monolayers of semiconducting transition dichalcogenides\cite{piatti_multi-valley_2018,zhang_band_2019}, and on diamond films\cite{piatti_orientation-dependent_2020} have led to the observation of interesting phenomena. These results, together with the spectroscopic capabilities discussed in this paper, indicate that a detailed quantitative analysis of the behaviour of ionic liquid gated devices has more surprises to offer.\\


\begin{thebibliography}{100}
	\urlstyle{rm}
	\expandafter\ifx\csname url\endcsname\relax
	\def\url#1{\texttt{#1}}\fi
	\expandafter\ifx\csname urlprefix\endcsname\relax\def\urlprefix{URL }\fi
	\expandafter\ifx\csname doiprefix\endcsname\relax\def\doiprefix{DOI: }\fi
	\providecommand{\bibinfo}[2]{#2}
	\providecommand{\eprint}[2][]{\url{#2}}
	
	\bibitem{ahn_electrostatic_2006}
	\bibinfo{author}{Ahn, C.~H.} \emph{et~al.}
	\newblock \bibinfo{journal}{\bibinfo{title}{Electrostatic modification of novel
			materials}}.
	\newblock {\emph{\JournalTitle{Reviews of Modern Physics}}}
	\textbf{\bibinfo{volume}{78}}, \bibinfo{pages}{1185--1212},
	\doiprefix\url{10.1103/RevModPhys.78.1185} (\bibinfo{year}{2006}).
	\newblock \bibinfo{note}{Publisher: American Physical Society}.
	
	\bibitem{sze_physics_2006}
	\bibinfo{author}{Sze, S.~M.} \& \bibinfo{author}{Ng, K.~K.}
	\newblock \emph{\bibinfo{title}{Physics of {Semiconductor} {Devices}}}
	(\bibinfo{publisher}{John Wiley \& Sons}, \bibinfo{year}{2006}).
	
	\bibitem{brattain_experiments_1955}
	\bibinfo{author}{Brattain, W.~H.} \& \bibinfo{author}{Garrett, C. G.~B.}
	\newblock \bibinfo{journal}{\bibinfo{title}{Experiments on the {Interface}
			between {Germanium} and an {Electrolyte}}}.
	\newblock {\emph{\JournalTitle{Bell System Technical Journal}}}
	\textbf{\bibinfo{volume}{34}}, \bibinfo{pages}{129--176},
	\doiprefix\url{https://doi.org/10.1002/j.1538-7305.1955.tb03766.x}
	(\bibinfo{year}{1955}).
	\newblock \bibinfo{note}{\_eprint:
		https://onlinelibrary.wiley.com/doi/pdf/10.1002/j.1538-7305.1955.tb03766.x}.
	
	\bibitem{bergveld_development_1970}
	\bibinfo{author}{Bergveld, P.}
	\newblock \bibinfo{journal}{\bibinfo{title}{Development of an {Ion}-{Sensitive}
			{Solid}-{State} {Device} for {Neurophysiological} {Measurements}}}.
	\newblock {\emph{\JournalTitle{IEEE Transactions on Biomedical Engineering}}}
	\textbf{\bibinfo{volume}{BME-17}}, \bibinfo{pages}{70--71},
	\doiprefix\url{10.1109/TBME.1970.4502688} (\bibinfo{year}{1970}).
	\newblock \bibinfo{note}{Conference Name: IEEE Transactions on Biomedical
		Engineering}.
	
	\bibitem{kruger_electrochemical_2001}
	\bibinfo{author}{Krüger, M.}, \bibinfo{author}{Buitelaar, M.~R.},
	\bibinfo{author}{Nussbaumer, T.}, \bibinfo{author}{Schönenberger, C.} \&
	\bibinfo{author}{Forró, L.}
	\newblock \bibinfo{journal}{\bibinfo{title}{Electrochemical carbon nanotube
			field-effect transistor}}.
	\newblock {\emph{\JournalTitle{Applied Physics Letters}}}
	\textbf{\bibinfo{volume}{78}}, \bibinfo{pages}{1291--1293},
	\doiprefix\url{10.1063/1.1350427} (\bibinfo{year}{2001}).
	\newblock \bibinfo{note}{Publisher: American Institute of Physics}.
	
	\bibitem{nilsson_bi-stable_2002}
	\bibinfo{author}{Nilsson, D.} \emph{et~al.}
	\newblock \bibinfo{journal}{\bibinfo{title}{Bi-stable and {Dynamic} {Current}
			{Modulation} in {Electrochemical} {Organic} {Transistors}}}.
	\newblock {\emph{\JournalTitle{Advanced Materials}}}
	\textbf{\bibinfo{volume}{14}}, \bibinfo{pages}{51--54},
	\doiprefix\url{https://doi.org/10.1002/1521-4095(20020104)14:1<51::AID-ADMA51>3.0.CO;2-#}
	(\bibinfo{year}{2002}).
	\newblock \bibinfo{note}{\_eprint:
		https://onlinelibrary.wiley.com/doi/pdf/10.1002/1521-4095\%2820020104\%2914\%3A1\%3C51\%3A\%3AAID-ADMA51\%3E3.0.CO\%3B2-\%23}.
	
	\bibitem{rosenblatt_high_2002}
	\bibinfo{author}{Rosenblatt, S.} \emph{et~al.}
	\newblock \bibinfo{journal}{\bibinfo{title}{High {Performance} {Electrolyte}
			{Gated} {Carbon} {Nanotube} {Transistors}}}.
	\newblock {\emph{\JournalTitle{Nano Letters}}} \textbf{\bibinfo{volume}{2}},
	\bibinfo{pages}{869--872}, \doiprefix\url{10.1021/nl025639a}
	(\bibinfo{year}{2002}).
	\newblock \bibinfo{note}{Publisher: American Chemical Society}.
	
	\bibitem{panzer_low-voltage_2005}
	\bibinfo{author}{Panzer, M.~J.}, \bibinfo{author}{Newman, C.~R.} \&
	\bibinfo{author}{Frisbie, C.~D.}
	\newblock \bibinfo{journal}{\bibinfo{title}{Low-voltage operation of a
			pentacene field-effect transistor with a polymer electrolyte gate
			dielectric}}.
	\newblock {\emph{\JournalTitle{Applied Physics Letters}}}
	\textbf{\bibinfo{volume}{86}}, \bibinfo{pages}{103503},
	\doiprefix\url{10.1063/1.1880434} (\bibinfo{year}{2005}).
	\newblock \bibinfo{note}{Publisher: American Institute of Physics}.
	
	\bibitem{shimotani_direct_2005}
	\bibinfo{author}{Shimotani, H.}, \bibinfo{author}{Diguet, G.} \&
	\bibinfo{author}{Iwasa, Y.}
	\newblock \bibinfo{journal}{\bibinfo{title}{Direct comparison of field-effect
			and electrochemical doping in regioregular poly(3-hexylthiophene)}}.
	\newblock {\emph{\JournalTitle{Applied Physics Letters}}}
	\textbf{\bibinfo{volume}{86}}, \bibinfo{pages}{022104},
	\doiprefix\url{10.1063/1.1850614} (\bibinfo{year}{2005}).
	\newblock \bibinfo{note}{Publisher: American Institute of Physics}.
	
	\bibitem{misra_electric_2007}
	\bibinfo{author}{Misra, R.}, \bibinfo{author}{McCarthy, M.} \&
	\bibinfo{author}{Hebard, A.~F.}
	\newblock \bibinfo{journal}{\bibinfo{title}{Electric field gating with ionic
			liquids}}.
	\newblock {\emph{\JournalTitle{Appl. Phys. Lett.}}}
	\textbf{\bibinfo{volume}{90}}, \bibinfo{pages}{2005--2008},
	\doiprefix\url{10.1063/1.2437663} (\bibinfo{year}{2007}).
	
	\bibitem{fujimoto_electric-double-layer_2013}
	\bibinfo{author}{Fujimoto, T.} \& \bibinfo{author}{Awaga, K.}
	\newblock \bibinfo{journal}{\bibinfo{title}{Electric-double-layer field-effect
			transistors with ionic liquids}}.
	\newblock {\emph{\JournalTitle{Physical Chemistry Chemical Physics}}}
	\textbf{\bibinfo{volume}{15}}, \bibinfo{pages}{8983--9006},
	\doiprefix\url{10.1039/C3CP50755F} (\bibinfo{year}{2013}).
	\newblock \bibinfo{note}{Publisher: The Royal Society of Chemistry}.
	
	\bibitem{kim_electrolyte-gated_2013}
	\bibinfo{author}{Kim, S.~H.} \emph{et~al.}
	\newblock \bibinfo{journal}{\bibinfo{title}{Electrolyte-{Gated} {Transistors}
			for {Organic} and {Printed} {Electronics}}}.
	\newblock {\emph{\JournalTitle{Advanced Materials}}}
	\textbf{\bibinfo{volume}{25}}, \bibinfo{pages}{1822--1846},
	\doiprefix\url{10.1002/adma.201202790} (\bibinfo{year}{2013}).
	\newblock \bibinfo{note}{\_eprint:
		https://onlinelibrary.wiley.com/doi/pdf/10.1002/adma.201202790}.
	
	\bibitem{leger_iontronics_2016}
	\bibinfo{author}{Leger, J.}, \bibinfo{author}{Berggren, M.},
	\bibinfo{author}{Carter, S.}, \bibinfo{author}{Berggren, M.} \&
	\bibinfo{author}{Carter, S.}
	\newblock \emph{\bibinfo{title}{Iontronics : {Ionic} {Carriers} in {Organic}
			{Electronic} {Materials} and {Devices}}} (\bibinfo{publisher}{CRC Press},
	\bibinfo{year}{2016}).
	
	\bibitem{bisri_endeavor_2017}
	\bibinfo{author}{Bisri, S.~Z.}, \bibinfo{author}{Shimizu, S.},
	\bibinfo{author}{Nakano, M.} \& \bibinfo{author}{Iwasa, Y.}
	\newblock \bibinfo{journal}{\bibinfo{title}{Endeavor of {Iontronics}: {From}
			{Fundamentals} to {Applications} of {Ion}-{Controlled} {Electronics}}}.
	\newblock {\emph{\JournalTitle{Advanced Materials}}}
	\textbf{\bibinfo{volume}{29}}, \bibinfo{pages}{1607054},
	\doiprefix\url{10.1002/adma.201607054} (\bibinfo{year}{2017}).
	\newblock \bibinfo{note}{\_eprint:
		https://www.onlinelibrary.wiley.com/doi/pdf/10.1002/adma.201607054}.
	
	\bibitem{ye_liquid-gated_2010}
	\bibinfo{author}{Ye, J.~T.} \emph{et~al.}
	\newblock \bibinfo{journal}{\bibinfo{title}{Liquid-gated interface
			superconductivity on an atomically flat film}}.
	\newblock {\emph{\JournalTitle{Nature Materials}}}
	\textbf{\bibinfo{volume}{9}}, \bibinfo{pages}{125--128},
	\doiprefix\url{10.1038/nmat2587} (\bibinfo{year}{2010}).
	\newblock \bibinfo{note}{Number: 2 Publisher: Nature Publishing Group}.
	
	\bibitem{ye_superconducting_2012}
	\bibinfo{author}{Ye, J.~T.} \emph{et~al.}
	\newblock \bibinfo{journal}{\bibinfo{title}{Superconducting {Dome} in a
			{Gate}-{Tuned} {Band} {Insulator}}}.
	\newblock {\emph{\JournalTitle{Science}}} \textbf{\bibinfo{volume}{338}},
	\bibinfo{pages}{1193--1196}, \doiprefix\url{10.1126/science.1228006}
	(\bibinfo{year}{2012}).
	\newblock \bibinfo{note}{Publisher: American Association for the Advancement of
		Science Section: Report}.
	
	\bibitem{jo_electrostatically_2015}
	\bibinfo{author}{Jo, S.}, \bibinfo{author}{Costanzo, D.},
	\bibinfo{author}{Berger, H.} \& \bibinfo{author}{Morpurgo, A.~F.}
	\newblock \bibinfo{journal}{\bibinfo{title}{Electrostatically {Induced}
			{Superconductivity} at the {Surface} of {WS2}}}.
	\newblock {\emph{\JournalTitle{Nano Letters}}} \textbf{\bibinfo{volume}{15}},
	\bibinfo{pages}{1197--1202}, \doiprefix\url{10.1021/nl504314c}
	(\bibinfo{year}{2015}).
	
	\bibitem{shi_superconductivity_2015}
	\bibinfo{author}{Shi, W.} \emph{et~al.}
	\newblock \bibinfo{journal}{\bibinfo{title}{Superconductivity {Series} in
			{Transition} {Metal} {Dichalcogenides} by {Ionic} {Gating}}}.
	\newblock {\emph{\JournalTitle{Scientific Reports}}}
	\textbf{\bibinfo{volume}{5}}, \bibinfo{pages}{1--10},
	\doiprefix\url{10.1038/srep12534} (\bibinfo{year}{2015}).
	
	\bibitem{biscaras_onset_2015}
	\bibinfo{author}{Biscaras, J.}, \bibinfo{author}{Chen, Z.},
	\bibinfo{author}{Paradisi, A.} \& \bibinfo{author}{Shukla, A.}
	\newblock \bibinfo{journal}{\bibinfo{title}{Onset of two-dimensional
			superconductivity in space charge doped few-layer molybdenum disulfide}}.
	\newblock {\emph{\JournalTitle{Nature Communications}}}
	\textbf{\bibinfo{volume}{6}}, \bibinfo{pages}{8826},
	\doiprefix\url{10.1038/ncomms9826} (\bibinfo{year}{2015}).
	\newblock \bibinfo{note}{Number: 1 Publisher: Nature Publishing Group}.
	
	\bibitem{costanzo_gate-induced_2016}
	\bibinfo{author}{Costanzo, D.}, \bibinfo{author}{Jo, S.},
	\bibinfo{author}{Berger, H.} \& \bibinfo{author}{Morpurgo, A.~F.}
	\newblock \bibinfo{journal}{\bibinfo{title}{Gate-induced superconductivity in
			atomically thin {MoS2} crystals}}.
	\newblock {\emph{\JournalTitle{Nature Nanotechnology}}}
	\textbf{\bibinfo{volume}{11}}, \bibinfo{pages}{339--344},
	\doiprefix\url{10.1038/nnano.2015.314} (\bibinfo{year}{2016}).
	
	\bibitem{costanzo_tunnelling_2018}
	\bibinfo{author}{Costanzo, D.}, \bibinfo{author}{Zhang, H.},
	\bibinfo{author}{Reddy, B.~A.}, \bibinfo{author}{Berger, H.} \&
	\bibinfo{author}{Morpurgo, A.~F.}
	\newblock \bibinfo{journal}{\bibinfo{title}{Tunnelling spectroscopy of
			gate-induced superconductivity in {MoS} 2}}.
	\newblock {\emph{\JournalTitle{Nature Nanotechnology}}}
	\textbf{\bibinfo{volume}{13}}, \bibinfo{pages}{483--488},
	\doiprefix\url{10.1038/s41565-018-0122-2} (\bibinfo{year}{2018}).
	\newblock \bibinfo{note}{Number: 6 Publisher: Nature Publishing Group}.
	
	\bibitem{lu_full_2018}
	\bibinfo{author}{Lu, J.} \emph{et~al.}
	\newblock \bibinfo{journal}{\bibinfo{title}{Full superconducting dome of strong
			{Ising} protection in gated monolayer {WS2}}}.
	\newblock {\emph{\JournalTitle{Proceedings of the National Academy of
				Sciences}}} \textbf{\bibinfo{volume}{115}}, \bibinfo{pages}{3551--3556},
	\doiprefix\url{10.1073/pnas.1716781115} (\bibinfo{year}{2018}).
	\newblock \bibinfo{note}{Publisher: National Academy of Sciences Section:
		Physical Sciences}.
	
	\bibitem{piatti_multi-valley_2018}
	\bibinfo{author}{Piatti, E.} \emph{et~al.}
	\newblock \bibinfo{journal}{\bibinfo{title}{Multi-{Valley} {Superconductivity}
			in {Ion}-{Gated} {MoS2} {Layers}}}.
	\newblock {\emph{\JournalTitle{Nano Letters}}} \textbf{\bibinfo{volume}{18}},
	\bibinfo{pages}{4821--4830}, \doiprefix\url{10.1021/acs.nanolett.8b01390}
	(\bibinfo{year}{2018}).
	\newblock \bibinfo{note}{Publisher: American Chemical Society}.
	
	\bibitem{kouno_superconductivity_2018}
	\bibinfo{author}{Kouno, S.} \emph{et~al.}
	\newblock \bibinfo{journal}{\bibinfo{title}{Superconductivity at 38 {K} at an
			electrochemical interface between an ionic liquid and {FeSe} 0.8 {Te} 0.2 on
			various substrates}}.
	\newblock {\emph{\JournalTitle{Scientific Reports}}}
	\textbf{\bibinfo{volume}{8}}, \bibinfo{pages}{14731},
	\doiprefix\url{10.1038/s41598-018-33121-7} (\bibinfo{year}{2018}).
	\newblock \bibinfo{note}{Number: 1 Publisher: Nature Publishing Group}.
	
	\bibitem{zeng_gate-induced_2018}
	\bibinfo{author}{Zeng, J.} \emph{et~al.}
	\newblock \bibinfo{journal}{\bibinfo{title}{Gate-{Induced} {Interfacial}
			{Superconductivity} in {1T}-{SnSe2}}}.
	\newblock {\emph{\JournalTitle{Nano Letters}}} \textbf{\bibinfo{volume}{18}},
	\bibinfo{pages}{1410--1415}, \doiprefix\url{10.1021/acs.nanolett.7b05157}
	(\bibinfo{year}{2018}).
	\newblock \bibinfo{note}{Publisher: American Chemical Society}.
	
	\bibitem{piatti_ambipolar_2019}
	\bibinfo{author}{Piatti, E.} \emph{et~al.}
	\newblock \bibinfo{journal}{\bibinfo{title}{Ambipolar suppression of
			superconductivity by ionic gating in optimally doped
			\$\{{\textbackslash}mathrm\{{BaFe}\}\}\_\{2\}\{({\textbackslash}mathrm\{{As}\},{\textbackslash}mathrm\{{P}\})\}\_\{2\}\$
			ultrathin films}}.
	\newblock {\emph{\JournalTitle{Physical Review Materials}}}
	\textbf{\bibinfo{volume}{3}}, \bibinfo{pages}{044801},
	\doiprefix\url{10.1103/PhysRevMaterials.3.044801} (\bibinfo{year}{2019}).
	\newblock \bibinfo{note}{Publisher: American Physical Society}.
	
	\bibitem{luryi_quantum_1988}
	\bibinfo{author}{Luryi, S.}
	\newblock \bibinfo{journal}{\bibinfo{title}{Quantum capacitance devices}}.
	\newblock {\emph{\JournalTitle{Applied Physics Letters}}}
	\textbf{\bibinfo{volume}{52}}, \bibinfo{pages}{501--503},
	\doiprefix\url{10.1063/1.99649} (\bibinfo{year}{1988}).
	\newblock \bibinfo{note}{Publisher: American Institute of Physics}.
	
	\bibitem{davies_physics_1997}
	\bibinfo{author}{Davies, J.~H.}
	\newblock \bibinfo{title}{The {Physics} of {Low}-dimensional {Semiconductors}:
		{An} {Introduction}}, \doiprefix\url{10.1017/CBO9780511819070}
	(\bibinfo{year}{1997}).
	
	\bibitem{ilani_measurement_2006}
	\bibinfo{author}{Ilani, S.}, \bibinfo{author}{Donev, L. a.~K.},
	\bibinfo{author}{Kindermann, M.} \& \bibinfo{author}{McEuen, P.~L.}
	\newblock \bibinfo{journal}{\bibinfo{title}{Measurement of the quantum
			capacitance of interacting electrons in carbon nanotubes}}.
	\newblock {\emph{\JournalTitle{Nature Physics}}} \textbf{\bibinfo{volume}{2}},
	\bibinfo{pages}{687--691}, \doiprefix\url{10.1038/nphys412}
	(\bibinfo{year}{2006}).
	\newblock \bibinfo{note}{Number: 10 Publisher: Nature Publishing Group}.
	
	\bibitem{xia_measurement_2009}
	\bibinfo{author}{Xia, J.}, \bibinfo{author}{Chen, F.}, \bibinfo{author}{Li, J.}
	\& \bibinfo{author}{Tao, N.}
	\newblock \bibinfo{journal}{\bibinfo{title}{Measurement of the quantum
			capacitance of graphene}}.
	\newblock {\emph{\JournalTitle{Nature Nanotechnology}}}
	\textbf{\bibinfo{volume}{4}}, \bibinfo{pages}{505--509},
	\doiprefix\url{10.1038/nnano.2009.177} (\bibinfo{year}{2009}).
	\newblock \bibinfo{note}{Number: 8 Publisher: Nature Publishing Group}.
	
	\bibitem{ihn_semiconductor_2010}
	\bibinfo{author}{Ihn, T.}
	\newblock \emph{\bibinfo{title}{Semiconductor {Nanostructures}: {Quantum}
			{States} and {Electronic} {Transport}}} (\bibinfo{publisher}{Oxford
		University Press}, \bibinfo{year}{2010}).
	\newblock \bibinfo{note}{Google-Books-ID: PAtqPgAACAAJ}.
	
	\bibitem{braga_quantitative_2012}
	\bibinfo{author}{Braga, D.}, \bibinfo{author}{Gutiérrez~Lezama, I.},
	\bibinfo{author}{Berger, H.} \& \bibinfo{author}{Morpurgo, A.~F.}
	\newblock \bibinfo{journal}{\bibinfo{title}{Quantitative {Determination} of the
			{Band} {Gap} of {WS2} with {Ambipolar} {Ionic} {Liquid}-{Gated}
			{Transistors}}}.
	\newblock {\emph{\JournalTitle{Nano Letters}}} \textbf{\bibinfo{volume}{12}},
	\bibinfo{pages}{5218--5223}, \doiprefix\url{10.1021/nl302389d}
	(\bibinfo{year}{2012}).
	
	\bibitem{zhang_ambipolar_2012}
	\bibinfo{author}{Zhang, Y.}, \bibinfo{author}{Ye, J.},
	\bibinfo{author}{Matsuhashi, Y.} \& \bibinfo{author}{Iwasa, Y.}
	\newblock \bibinfo{journal}{\bibinfo{title}{Ambipolar {MoS2} {Thin} {Flake}
			{Transistors}}}.
	\newblock {\emph{\JournalTitle{Nano Letters}}} \textbf{\bibinfo{volume}{12}},
	\bibinfo{pages}{1136--1140}, \doiprefix\url{10.1021/nl2021575}
	(\bibinfo{year}{2012}).
	\newblock \bibinfo{note}{Publisher: American Chemical Society}.
	
	\bibitem{ubrig_scanning_2014}
	\bibinfo{author}{Ubrig, N.}, \bibinfo{author}{Jo, S.}, \bibinfo{author}{Berger,
		H.}, \bibinfo{author}{Morpurgo, A.~F.} \& \bibinfo{author}{Kuzmenko, A.~B.}
	\newblock \bibinfo{journal}{\bibinfo{title}{Scanning photocurrent microscopy
			reveals electron-hole asymmetry in ionic liquid-gated {WS2} transistors}}.
	\newblock {\emph{\JournalTitle{Applied Physics Letters}}}
	\textbf{\bibinfo{volume}{104}}, \bibinfo{pages}{171112},
	\doiprefix\url{10.1063/1.4872002} (\bibinfo{year}{2014}).
	
	\bibitem{ovchinnikov_disorder_2016}
	\bibinfo{author}{Ovchinnikov, D.} \emph{et~al.}
	\newblock \bibinfo{journal}{\bibinfo{title}{Disorder engineering and
			conductivity dome in {ReS} 2 with electrolyte gating}}.
	\newblock {\emph{\JournalTitle{Nature Communications}}}
	\textbf{\bibinfo{volume}{7}}, \bibinfo{pages}{12391},
	\doiprefix\url{10.1038/ncomms12391} (\bibinfo{year}{2016}).
	\newblock \bibinfo{note}{Number: 1 Publisher: Nature Publishing Group}.
	
	\bibitem{zhang_electrically_2014}
	\bibinfo{author}{Zhang, Y.~J.}, \bibinfo{author}{Oka, T.},
	\bibinfo{author}{Suzuki, R.}, \bibinfo{author}{Ye, J.~T.} \&
	\bibinfo{author}{Iwasa, Y.}
	\newblock \bibinfo{journal}{\bibinfo{title}{Electrically {Switchable} {Chiral}
			{Light}-{Emitting} {Transistor}}}.
	\newblock {\emph{\JournalTitle{Science}}} \textbf{\bibinfo{volume}{344}},
	\bibinfo{pages}{725--728}, \doiprefix\url{10.1126/science.1251329}
	(\bibinfo{year}{2014}).
	
	\bibitem{jo_mono-_2014}
	\bibinfo{author}{Jo, S.}, \bibinfo{author}{Ubrig, N.}, \bibinfo{author}{Berger,
		H.}, \bibinfo{author}{Kuzmenko, A.~B.} \& \bibinfo{author}{Morpurgo, A.~F.}
	\newblock \bibinfo{journal}{\bibinfo{title}{Mono- and {Bilayer} {WS2}
			{Light}-{Emitting} {Transistors}}}.
	\newblock {\emph{\JournalTitle{Nano Letters}}} \textbf{\bibinfo{volume}{14}},
	\bibinfo{pages}{2019--2025}, \doiprefix\url{10.1021/nl500171v}
	(\bibinfo{year}{2014}).
	\newblock \bibinfo{note}{Publisher: American Chemical Society}.
	
	\bibitem{ponomarev_ambipolar_2015}
	\bibinfo{author}{Ponomarev, E.}, \bibinfo{author}{Gutierrez-Lezama, I.},
	\bibinfo{author}{Ubrig, N.} \& \bibinfo{author}{Morpurgo, A.~F.}
	\newblock \bibinfo{journal}{\bibinfo{title}{Ambipolar {Light}-{Emitting}
			{Transistors} on {Chemical} {Vapor} {Deposited} {Monolayer} {MoS2}}}.
	\newblock {\emph{\JournalTitle{Nano Letters}}} \textbf{\bibinfo{volume}{15}},
	\bibinfo{pages}{8289--8294}, \doiprefix\url{10.1021/acs.nanolett.5b03885}
	(\bibinfo{year}{2015}).
	
	\bibitem{gutierrez-lezama_electroluminescence_2016}
	\bibinfo{author}{Gutiérrez-Lezama, I.}, \bibinfo{author}{Reddy, B.~A.},
	\bibinfo{author}{Ubrig, N.} \& \bibinfo{author}{Morpurgo, A.~F.}
	\newblock \bibinfo{journal}{\bibinfo{title}{Electroluminescence from indirect
			band gap semiconductor {ReS} 2}}.
	\newblock {\emph{\JournalTitle{2D Materials}}} \textbf{\bibinfo{volume}{3}},
	\bibinfo{pages}{045016}, \doiprefix\url{10.1088/2053-1583/3/4/045016}
	(\bibinfo{year}{2016}).
	
	\bibitem{lezama_surface_2014}
	\bibinfo{author}{Lezama, I.~G.} \emph{et~al.}
	\newblock \bibinfo{journal}{\bibinfo{title}{Surface transport and band gap
			structure of exfoliated {2H}-{MoTe} 2 crystals}}.
	\newblock {\emph{\JournalTitle{2D Materials}}} \textbf{\bibinfo{volume}{1}},
	\bibinfo{pages}{021002}, \doiprefix\url{10.1088/2053-1583/1/2/021002}
	(\bibinfo{year}{2014}).
	\newblock \bibinfo{note}{Publisher: IOP Publishing}.
	
	\bibitem{ponomarev_semiconducting_2018}
	\bibinfo{author}{Ponomarev, E.}, \bibinfo{author}{Ubrig, N.},
	\bibinfo{author}{Gutiérrez-Lezama, I.}, \bibinfo{author}{Berger, H.} \&
	\bibinfo{author}{Morpurgo, A.~F.}
	\newblock \bibinfo{journal}{\bibinfo{title}{Semiconducting van der {Waals}
			{Interfaces} as {Artificial} {Semiconductors}}}.
	\newblock {\emph{\JournalTitle{Nano Letters}}} \textbf{\bibinfo{volume}{18}},
	\bibinfo{pages}{5146--5152}, \doiprefix\url{10.1021/acs.nanolett.8b02066}
	(\bibinfo{year}{2018}).
	\newblock \bibinfo{note}{Publisher: American Chemical Society}.
	
	\bibitem{saito_ambipolar_2015}
	\bibinfo{author}{Saito, Y.} \& \bibinfo{author}{Iwasa, Y.}
	\newblock \bibinfo{journal}{\bibinfo{title}{Ambipolar {Insulator}-to-{Metal}
			{Transition} in {Black} {Phosphorus} by {Ionic}-{Liquid} {Gating}}}.
	\newblock {\emph{\JournalTitle{ACS Nano}}} \textbf{\bibinfo{volume}{9}},
	\bibinfo{pages}{3192--3198}, \doiprefix\url{10.1021/acsnano.5b00497}
	(\bibinfo{year}{2015}).
	
	\bibitem{ciarrocchi_thickness-modulated_2018}
	\bibinfo{author}{Ciarrocchi, A.}, \bibinfo{author}{Avsar, A.},
	\bibinfo{author}{Ovchinnikov, D.} \& \bibinfo{author}{Kis, A.}
	\newblock \bibinfo{journal}{\bibinfo{title}{Thickness-modulated
			metal-to-semiconductor transformation in a transition metal dichalcogenide}}.
	\newblock {\emph{\JournalTitle{Nature Communications}}}
	\textbf{\bibinfo{volume}{9}}, \bibinfo{pages}{919},
	\doiprefix\url{10.1038/s41467-018-03436-0} (\bibinfo{year}{2018}).
	
	\bibitem{zhang_band_2019}
	\bibinfo{author}{Zhang, H.}, \bibinfo{author}{Berthod, C.},
	\bibinfo{author}{Berger, H.}, \bibinfo{author}{Giamarchi, T.} \&
	\bibinfo{author}{Morpurgo, A.~F.}
	\newblock \bibinfo{journal}{\bibinfo{title}{Band {Filling} and {Cross}
			{Quantum} {Capacitance} in {Ion}-{Gated} {Semiconducting} {Transition}
			{Metal} {Dichalcogenide} {Monolayers}}}.
	\newblock {\emph{\JournalTitle{Nano Letters}}} \textbf{\bibinfo{volume}{19}},
	\bibinfo{pages}{8836--8845}, \doiprefix\url{10.1021/acs.nanolett.9b03667}
	(\bibinfo{year}{2019}).
	\newblock \bibinfo{note}{Publisher: American Chemical Society}.
	
	\bibitem{reddy_synthetic_2020}
	\bibinfo{author}{Reddy, B.~A.} \emph{et~al.}
	\newblock \bibinfo{journal}{\bibinfo{title}{Synthetic {Semimetals} with van der
			{Waals} {Interfaces}}}.
	\newblock {\emph{\JournalTitle{Nano Letters}}} \textbf{\bibinfo{volume}{20}},
	\bibinfo{pages}{1322--1328}, \doiprefix\url{10.1021/acs.nanolett.9b04810}
	(\bibinfo{year}{2020}).
	
	\bibitem{piatti_orientation-dependent_2020}
	\bibinfo{author}{Piatti, E.}, \bibinfo{author}{Pasquarelli, A.} \&
	\bibinfo{author}{Gonnelli, R.~S.}
	\newblock \bibinfo{journal}{\bibinfo{title}{Orientation-dependent electric
			transport and band filling in hole co-doped epitaxial diamond films}}.
	\newblock {\emph{\JournalTitle{Applied Surface Science}}}
	\textbf{\bibinfo{volume}{528}}, \bibinfo{pages}{146795},
	\doiprefix\url{10.1016/j.apsusc.2020.146795} (\bibinfo{year}{2020}).
	
	\bibitem{bardeen_john_semiconductor_1956}
	\bibinfo{author}{Bardeen, J.}
	\newblock \bibinfo{title}{Semiconductor research leading to the point contact
		transistor - {Nobel} {Lectures} in {Physics} 1942 – 1962}
	(\bibinfo{year}{1956}).
	
	\bibitem{ono_comparative_2009}
	\bibinfo{author}{Ono, S.}, \bibinfo{author}{Miwa, K.}, \bibinfo{author}{Seki,
		S.} \& \bibinfo{author}{Takeya, J.}
	\newblock \bibinfo{journal}{\bibinfo{title}{A comparative study of organic
			single-crystal transistors gated with various ionic-liquid electrolytes}}.
	\newblock {\emph{\JournalTitle{Applied Physics Letters}}}
	\textbf{\bibinfo{volume}{94}}, \bibinfo{pages}{063301},
	\doiprefix\url{10.1063/1.3079401} (\bibinfo{year}{2009}).
	\newblock \bibinfo{note}{Publisher: American Institute of Physics}.
	
	\bibitem{novoselov_two-dimensional_2005}
	\bibinfo{author}{Novoselov, K.~S.} \emph{et~al.}
	\newblock \bibinfo{journal}{\bibinfo{title}{Two-{Dimensional} {Atomic}
			{Crystals}}}.
	\newblock {\emph{\JournalTitle{Proc. Natl. Acad. Sci. U. S. A.}}}
	\textbf{\bibinfo{volume}{102}}, \bibinfo{pages}{10451--10453},
	\doiprefix\url{10.1073/pnas.0502848102} (\bibinfo{year}{2005}).
	
	\bibitem{geim_rise_2007}
	\bibinfo{author}{Geim, A.~K.} \& \bibinfo{author}{Novoselov, K.~S.}
	\newblock \bibinfo{journal}{\bibinfo{title}{The rise of graphene}}.
	\newblock {\emph{\JournalTitle{Nature Materials}}}
	\textbf{\bibinfo{volume}{6}}, \bibinfo{pages}{183--191},
	\doiprefix\url{10.1038/nmat1849} (\bibinfo{year}{2007}).
	\newblock \bibinfo{note}{Number: 3 Publisher: Nature Publishing Group}.
	
	\bibitem{yin_ferroelectric-induced_2017}
	\bibinfo{author}{Yin, L.} \emph{et~al.}
	\newblock \bibinfo{journal}{\bibinfo{title}{Ferroelectric-induced carrier
			modulation for ambipolar transition metal dichalcogenide transistors}}.
	\newblock {\emph{\JournalTitle{Applied Physics Letters}}}
	\textbf{\bibinfo{volume}{110}}, \bibinfo{pages}{123106},
	\doiprefix\url{10.1063/1.4979088} (\bibinfo{year}{2017}).
	\newblock \bibinfo{note}{Publisher: American Institute of Physics}.
	
	\bibitem{seo_low-frequency_2019}
	\bibinfo{author}{Seo, S.~G.}, \bibinfo{author}{Hong, J.~H.},
	\bibinfo{author}{Ryu, J.~H.} \& \bibinfo{author}{Jin, S.~H.}
	\newblock \bibinfo{journal}{\bibinfo{title}{Low-{Frequency} {Noise}
			{Characteristics} in {Multilayer} {MoTe2} {FETs} {With} {Hydrophobic}
			{Amorphous} {Fluoropolymers}}}.
	\newblock {\emph{\JournalTitle{IEEE Electron Device Letters}}}
	\textbf{\bibinfo{volume}{40}}, \bibinfo{pages}{251--254},
	\doiprefix\url{10.1109/LED.2018.2889904} (\bibinfo{year}{2019}).
	\newblock \bibinfo{note}{Conference Name: IEEE Electron Device Letters}.
	
	\bibitem{xia_correlation_2009}
	\bibinfo{author}{Xia, Y.}, \bibinfo{author}{Cho, J.}, \bibinfo{author}{Paulsen,
		B.}, \bibinfo{author}{Frisbie, C.~D.} \& \bibinfo{author}{Renn, M.~J.}
	\newblock \bibinfo{journal}{\bibinfo{title}{Correlation of on-state conductance
			with referenced electrochemical potential in ion gel gated polymer
			transistors}}.
	\newblock {\emph{\JournalTitle{Applied Physics Letters}}}
	\textbf{\bibinfo{volume}{94}}, \bibinfo{pages}{013304},
	\doiprefix\url{10.1063/1.3058694} (\bibinfo{year}{2009}).
	\newblock \bibinfo{note}{Publisher: American Institute of Physics}.
	
	\bibitem{ueno_electric-field-induced_2008}
	\bibinfo{author}{Ueno, K.} \emph{et~al.}
	\newblock \bibinfo{journal}{\bibinfo{title}{Electric-field-induced
			superconductivity in an insulator}}.
	\newblock {\emph{\JournalTitle{Nature Materials}}}
	\textbf{\bibinfo{volume}{7}}, \bibinfo{pages}{855--858},
	\doiprefix\url{10.1038/nmat2298} (\bibinfo{year}{2008}).
	\newblock \bibinfo{note}{Number: 11 Publisher: Nature Publishing Group}.
	
	\bibitem{yamada_electrically_2011}
	\bibinfo{author}{Yamada, Y.} \emph{et~al.}
	\newblock \bibinfo{journal}{\bibinfo{title}{Electrically {Induced}
			{Ferromagnetism} at {Room} {Temperature} in {Cobalt}-{Doped} {Titanium}
			{Dioxide}}}.
	\newblock {\emph{\JournalTitle{Science}}} \textbf{\bibinfo{volume}{332}},
	\bibinfo{pages}{1065--1067}, \doiprefix\url{10.1126/science.1202152}
	(\bibinfo{year}{2011}).
	\newblock \bibinfo{note}{Publisher: American Association for the Advancement of
		Science Section: Report}.
	
	\bibitem{ueno_discovery_2011}
	\bibinfo{author}{Ueno, K.} \emph{et~al.}
	\newblock \bibinfo{journal}{\bibinfo{title}{Discovery of superconductivity in
			{KTaO} 3 by electrostatic carrier doping}}.
	\newblock {\emph{\JournalTitle{Nature Nanotechnology}}}
	\textbf{\bibinfo{volume}{6}}, \bibinfo{pages}{408--412},
	\doiprefix\url{10.1038/nnano.2011.78} (\bibinfo{year}{2011}).
	\newblock \bibinfo{note}{Number: 7 Publisher: Nature Publishing Group}.
	
	\bibitem{sohier_enhanced_2019}
	\bibinfo{author}{Sohier, T.} \emph{et~al.}
	\newblock \bibinfo{journal}{\bibinfo{title}{Enhanced {Electron}-{Phonon}
			{Interaction} in {Multivalley} {Materials}}}.
	\newblock {\emph{\JournalTitle{Physical Review X}}}
	\textbf{\bibinfo{volume}{9}}, \bibinfo{pages}{031019},
	\doiprefix\url{10.1103/PhysRevX.9.031019} (\bibinfo{year}{2019}).
	
	\bibitem{liang_electronic_2014}
	\bibinfo{author}{Liang, L.} \emph{et~al.}
	\newblock \bibinfo{journal}{\bibinfo{title}{Electronic {Bandgap} and {Edge}
			{Reconstruction} in {Phosphorene} {Materials}}}.
	\newblock {\emph{\JournalTitle{Nano Letters}}} \textbf{\bibinfo{volume}{14}},
	\bibinfo{pages}{6400--6406}, \doiprefix\url{10.1021/nl502892t}
	(\bibinfo{year}{2014}).
	\newblock \bibinfo{note}{Publisher: American Chemical Society}.
	
	\bibitem{jeong_suppression_2013}
	\bibinfo{author}{Jeong, J.} \emph{et~al.}
	\newblock \bibinfo{journal}{\bibinfo{title}{Suppression of {Metal}-{Insulator}
			{Transition} in {VO2} by {Electric} {Field}–{Induced} {Oxygen} {Vacancy}
			{Formation}}}.
	\newblock {\emph{\JournalTitle{Science}}} \textbf{\bibinfo{volume}{339}},
	\bibinfo{pages}{1402--1405}, \doiprefix\url{10.1126/science.1230512}
	(\bibinfo{year}{2013}).
	\newblock \bibinfo{note}{Publisher: American Association for the Advancement of
		Science Section: Report}.
	
	\bibitem{fete_ionic_2016}
	\bibinfo{author}{Fête, A.}, \bibinfo{author}{Rossi, L.},
	\bibinfo{author}{Augieri, A.} \& \bibinfo{author}{Senatore, C.}
	\newblock \bibinfo{journal}{\bibinfo{title}{Ionic liquid gating of ultra-thin
			{YBa2Cu3O7}-x films}}.
	\newblock {\emph{\JournalTitle{Applied Physics Letters}}}
	\textbf{\bibinfo{volume}{109}}, \bibinfo{pages}{192601},
	\doiprefix\url{10.1063/1.4967197} (\bibinfo{year}{2016}).
	\newblock \bibinfo{note}{Publisher: American Institute of Physics}.
	
	\bibitem{scherwitzl_electric-field_2010}
	\bibinfo{author}{Scherwitzl, R.} \emph{et~al.}
	\newblock \bibinfo{journal}{\bibinfo{title}{Electric-{Field} {Control} of the
			{Metal}-{Insulator} {Transition} in {Ultrathin} {NdNiO3} {Films}}}.
	\newblock {\emph{\JournalTitle{Advanced Materials}}}
	\textbf{\bibinfo{volume}{22}}, \bibinfo{pages}{5517--5520},
	\doiprefix\url{10.1002/adma.201003241} (\bibinfo{year}{2010}).
	\newblock \bibinfo{note}{\_eprint:
		https://onlinelibrary.wiley.com/doi/pdf/10.1002/adma.201003241}.
	
	\bibitem{wang_scattering_2020}
	\bibinfo{author}{Wang, H.} \emph{et~al.}
	\newblock \bibinfo{journal}{\bibinfo{title}{Scattering mechanisms and mobility
			enhancement in epitaxial {BaSnO3} thin films probed via electrolyte gating}}.
	\newblock {\emph{\JournalTitle{APL Materials}}} \textbf{\bibinfo{volume}{8}},
	\bibinfo{pages}{071113}, \doiprefix\url{10.1063/5.0017227}
	(\bibinfo{year}{2020}).
	\newblock \bibinfo{note}{Publisher: American Institute of Physics}.
	
	\bibitem{chernikov_exciton_2014}
	\bibinfo{author}{Chernikov, A.} \emph{et~al.}
	\newblock \bibinfo{journal}{\bibinfo{title}{Exciton {Binding} {Energy} and
			{Nonhydrogenic} {Rydberg} {Series} in {Monolayer}
			\$\{{\textbackslash}mathrm\{{WS}\}\}\_\{2\}\$}}.
	\newblock {\emph{\JournalTitle{Physical Review Letters}}}
	\textbf{\bibinfo{volume}{113}}, \bibinfo{pages}{076802},
	\doiprefix\url{10.1103/PhysRevLett.113.076802} (\bibinfo{year}{2014}).
	\newblock \bibinfo{note}{Publisher: American Physical Society}.
	
	\bibitem{ruppert_optical_2014}
	\bibinfo{author}{Ruppert, C.}, \bibinfo{author}{Aslan, O.~B.} \&
	\bibinfo{author}{Heinz, T.~F.}
	\newblock \bibinfo{journal}{\bibinfo{title}{Optical {Properties} and {Band}
			{Gap} of {Single}- and {Few}-{Layer} {MoTe2} {Crystals}}}.
	\newblock {\emph{\JournalTitle{Nano Letters}}} \textbf{\bibinfo{volume}{14}},
	\bibinfo{pages}{6231--6236}, \doiprefix\url{10.1021/nl502557g}
	(\bibinfo{year}{2014}).
	\newblock \bibinfo{note}{Publisher: American Chemical Society}.
	
	\bibitem{klots_probing_2014}
	\bibinfo{author}{Klots, A.~R.} \emph{et~al.}
	\newblock \bibinfo{journal}{\bibinfo{title}{Probing excitonic states in
			suspended two-dimensional semiconductors by photocurrent spectroscopy}}.
	\newblock {\emph{\JournalTitle{Scientific Reports}}}
	\textbf{\bibinfo{volume}{4}}, \bibinfo{pages}{6608},
	\doiprefix\url{10.1038/srep06608} (\bibinfo{year}{2014}).
	\newblock \bibinfo{note}{Number: 1 Publisher: Nature Publishing Group}.
	
	\bibitem{lezama_indirect--direct_2015}
	\bibinfo{author}{Lezama, I.~G.} \emph{et~al.}
	\newblock \bibinfo{journal}{\bibinfo{title}{Indirect-to-{Direct} {Band} {Gap}
			{Crossover} in {Few}-{Layer} {MoTe2}}}.
	\newblock {\emph{\JournalTitle{Nano Letters}}} \textbf{\bibinfo{volume}{15}},
	\bibinfo{pages}{2336--2342}, \doiprefix\url{10.1021/nl5045007}
	(\bibinfo{year}{2015}).
	\newblock \bibinfo{note}{Publisher: American Chemical Society}.
	
	\bibitem{hanbicki_measurement_2015}
	\bibinfo{author}{Hanbicki, A.~T.}, \bibinfo{author}{Currie, M.},
	\bibinfo{author}{Kioseoglou, G.}, \bibinfo{author}{Friedman, A.~L.} \&
	\bibinfo{author}{Jonker, B.~T.}
	\newblock \bibinfo{journal}{\bibinfo{title}{Measurement of high exciton binding
			energy in the monolayer transition-metal dichalcogenides {WS2} and {WSe2}}}.
	\newblock {\emph{\JournalTitle{Solid State Communications}}}
	\textbf{\bibinfo{volume}{203}}, \bibinfo{pages}{16--20},
	\doiprefix\url{10.1016/j.ssc.2014.11.005} (\bibinfo{year}{2015}).
	
	\bibitem{poellmann_resonant_2015}
	\bibinfo{author}{Poellmann, C.} \emph{et~al.}
	\newblock \bibinfo{journal}{\bibinfo{title}{Resonant internal quantum
			transitions and femtosecond radiative decay of excitons in monolayer {WSe}
			2}}.
	\newblock {\emph{\JournalTitle{Nature Materials}}}
	\textbf{\bibinfo{volume}{14}}, \bibinfo{pages}{889--893},
	\doiprefix\url{10.1038/nmat4356} (\bibinfo{year}{2015}).
	\newblock \bibinfo{note}{Number: 9 Publisher: Nature Publishing Group}.
	
	\bibitem{wang_giant_2015}
	\bibinfo{author}{Wang, G.} \emph{et~al.}
	\newblock \bibinfo{journal}{\bibinfo{title}{Giant {Enhancement} of the
			{Optical} {Second}-{Harmonic} {Emission} of
			\$\{{\textbackslash}mathrm\{{WSe}\}\}\_\{2\}\$ {Monolayers} by {Laser}
			{Excitation} at {Exciton} {Resonances}}}.
	\newblock {\emph{\JournalTitle{Physical Review Letters}}}
	\textbf{\bibinfo{volume}{114}}, \bibinfo{pages}{097403},
	\doiprefix\url{10.1103/PhysRevLett.114.097403} (\bibinfo{year}{2015}).
	\newblock \bibinfo{note}{Publisher: American Physical Society}.
	
	\bibitem{hill_observation_2015}
	\bibinfo{author}{Hill, H.~M.} \emph{et~al.}
	\newblock \bibinfo{journal}{\bibinfo{title}{Observation of {Excitonic}
			{Rydberg} {States} in {Monolayer} {MoS2} and {WS2} by {Photoluminescence}
			{Excitation} {Spectroscopy}}}.
	\newblock {\emph{\JournalTitle{Nano Letters}}} \textbf{\bibinfo{volume}{15}},
	\bibinfo{pages}{2992--2997}, \doiprefix\url{10.1021/nl504868p}
	(\bibinfo{year}{2015}).
	\newblock \bibinfo{note}{Publisher: American Chemical Society}.
	
	\bibitem{raja_coulomb_2017}
	\bibinfo{author}{Raja, A.} \emph{et~al.}
	\newblock \bibinfo{journal}{\bibinfo{title}{Coulomb engineering of the bandgap
			and excitons in two-dimensional materials}}.
	\newblock {\emph{\JournalTitle{Nature Communications}}}
	\textbf{\bibinfo{volume}{8}}, \bibinfo{pages}{15251},
	\doiprefix\url{10.1038/ncomms15251} (\bibinfo{year}{2017}).
	\newblock \bibinfo{note}{Number: 1 Publisher: Nature Publishing Group}.
	
	\bibitem{yore_large_2017}
	\bibinfo{author}{Yore, A.~E.} \emph{et~al.}
	\newblock \bibinfo{journal}{\bibinfo{title}{Large array fabrication of high
			performance monolayer {MoS2} photodetectors}}.
	\newblock {\emph{\JournalTitle{Applied Physics Letters}}}
	\textbf{\bibinfo{volume}{111}}, \bibinfo{pages}{043110},
	\doiprefix\url{10.1063/1.4995984} (\bibinfo{year}{2017}).
	\newblock \bibinfo{note}{Publisher: American Institute of Physics}.
	
	\bibitem{lu_bandgap_2014}
	\bibinfo{author}{Lu, C.-P.}, \bibinfo{author}{Li, G.}, \bibinfo{author}{Mao,
		J.}, \bibinfo{author}{Wang, L.-M.} \& \bibinfo{author}{Andrei, E.~Y.}
	\newblock \bibinfo{journal}{\bibinfo{title}{Bandgap, {Mid}-{Gap} {States}, and
			{Gating} {Effects} in {MoS2}}}.
	\newblock {\emph{\JournalTitle{Nano Letters}}} \textbf{\bibinfo{volume}{14}},
	\bibinfo{pages}{4628--4633}, \doiprefix\url{10.1021/nl501659n}
	(\bibinfo{year}{2014}).
	\newblock \bibinfo{note}{Publisher: American Chemical Society}.
	
	\bibitem{zhang_direct_2014}
	\bibinfo{author}{Zhang, C.}, \bibinfo{author}{Johnson, A.},
	\bibinfo{author}{Hsu, C.-L.}, \bibinfo{author}{Li, L.-J.} \&
	\bibinfo{author}{Shih, C.-K.}
	\newblock \bibinfo{journal}{\bibinfo{title}{Direct {Imaging} of {Band}
			{Profile} in {Single} {Layer} {MoS2} on {Graphite}: {Quasiparticle} {Energy}
			{Gap}, {Metallic} {Edge} {States}, and {Edge} {Band} {Bending}}}.
	\newblock {\emph{\JournalTitle{Nano Letters}}} \textbf{\bibinfo{volume}{14}},
	\bibinfo{pages}{2443--2447}, \doiprefix\url{10.1021/nl501133c}
	(\bibinfo{year}{2014}).
	\newblock \bibinfo{note}{Publisher: American Chemical Society}.
	
	\bibitem{chiu_determination_2015}
	\bibinfo{author}{Chiu, M.-H.} \emph{et~al.}
	\newblock \bibinfo{journal}{\bibinfo{title}{Determination of band alignment in
			the single-layer {MoS} 2 /{WSe} 2 heterojunction}}.
	\newblock {\emph{\JournalTitle{Nature Communications}}}
	\textbf{\bibinfo{volume}{6}}, \bibinfo{pages}{7666},
	\doiprefix\url{10.1038/ncomms8666} (\bibinfo{year}{2015}).
	\newblock \bibinfo{note}{Number: 1 Publisher: Nature Publishing Group}.
	
	\bibitem{huang_bandgap_2015}
	\bibinfo{author}{Huang, Y.~L.} \emph{et~al.}
	\newblock \bibinfo{journal}{\bibinfo{title}{Bandgap tunability at single-layer
			molybdenum disulphide grain boundaries}}.
	\newblock {\emph{\JournalTitle{Nature Communications}}}
	\textbf{\bibinfo{volume}{6}}, \bibinfo{pages}{6298},
	\doiprefix\url{10.1038/ncomms7298} (\bibinfo{year}{2015}).
	\newblock \bibinfo{note}{Number: 1 Publisher: Nature Publishing Group}.
	
	\bibitem{rigosi_electronic_2016}
	\bibinfo{author}{Rigosi, A.~F.}, \bibinfo{author}{Hill, H.~M.},
	\bibinfo{author}{Rim, K.~T.}, \bibinfo{author}{Flynn, G.~W.} \&
	\bibinfo{author}{Heinz, T.~F.}
	\newblock \bibinfo{journal}{\bibinfo{title}{Electronic band gaps and exciton
			binding energies in monolayer
			\${\textbackslash}mathrm\{{M}\}\{{\textbackslash}mathrm\{o\}\}\_\{x\}\{{\textbackslash}mathrm\{{W}\}\}\_\{1{\textbackslash}text\{{\textbackslash}ensuremath\{-\}\}x\}\{{\textbackslash}mathrm\{{S}\}\}\_\{2\}\$
			transition metal dichalcogenide alloys probed by scanning tunneling and
			optical spectroscopy}}.
	\newblock {\emph{\JournalTitle{Physical Review B}}}
	\textbf{\bibinfo{volume}{94}}, \bibinfo{pages}{075440},
	\doiprefix\url{10.1103/PhysRevB.94.075440} (\bibinfo{year}{2016}).
	\newblock \bibinfo{note}{Publisher: American Physical Society}.
	
	\bibitem{hill_band_2016}
	\bibinfo{author}{Hill, H.~M.}, \bibinfo{author}{Rigosi, A.~F.},
	\bibinfo{author}{Rim, K.~T.}, \bibinfo{author}{Flynn, G.~W.} \&
	\bibinfo{author}{Heinz, T.~F.}
	\newblock \bibinfo{journal}{\bibinfo{title}{Band {Alignment} in {MoS2}/{WS2}
			{Transition} {Metal} {Dichalcogenide} {Heterostructures} {Probed} by
			{Scanning} {Tunneling} {Microscopy} and {Spectroscopy}}}.
	\newblock {\emph{\JournalTitle{Nano Letters}}} \textbf{\bibinfo{volume}{16}},
	\bibinfo{pages}{4831--4837}, \doiprefix\url{10.1021/acs.nanolett.6b01007}
	(\bibinfo{year}{2016}).
	\newblock \bibinfo{note}{Publisher: American Chemical Society}.
	
	\bibitem{he_tightly_2014}
	\bibinfo{author}{He, K.} \emph{et~al.}
	\newblock \bibinfo{journal}{\bibinfo{title}{Tightly {Bound} {Excitons} in
			{Monolayer} \$\{{\textbackslash}mathrm\{{WSe}\}\}\_\{2\}\$}}.
	\newblock {\emph{\JournalTitle{Physical Review Letters}}}
	\textbf{\bibinfo{volume}{113}}, \bibinfo{pages}{026803},
	\doiprefix\url{10.1103/PhysRevLett.113.026803} (\bibinfo{year}{2014}).
	\newblock \bibinfo{note}{Publisher: American Physical Society}.
	
	\bibitem{ugeda_observation_2014}
	\bibinfo{author}{Ugeda, M.~M.} \emph{et~al.}
	\newblock \bibinfo{journal}{\bibinfo{title}{Observation of giant bandgap
			renormalization and excitonic effects in a monolayer transition metal
			dichalcogenide semiconductor}}.
	\newblock {\emph{\JournalTitle{Nature Materials}}}
	\textbf{\bibinfo{volume}{13}}, \bibinfo{pages}{1091--1095},
	\doiprefix\url{10.1038/nmat4061} (\bibinfo{year}{2014}).
	\newblock \bibinfo{note}{ArXiv: 1404.2331}.
	
	\bibitem{wang_colloquium_2018}
	\bibinfo{author}{Wang, G.} \emph{et~al.}
	\newblock \bibinfo{journal}{\bibinfo{title}{Colloquium : {Excitons} in
			atomically thin transition metal dichalcogenides}}.
	\newblock {\emph{\JournalTitle{Reviews of Modern Physics}}}
	\textbf{\bibinfo{volume}{90}}, \bibinfo{pages}{021001},
	\doiprefix\url{10.1103/RevModPhys.90.021001} (\bibinfo{year}{2018}).
	
	\bibitem{splendiani_emerging_2010}
	\bibinfo{author}{Splendiani, A.} \emph{et~al.}
	\newblock \bibinfo{journal}{\bibinfo{title}{Emerging {Photoluminescence} in
			{Monolayer} {MoS2}}}.
	\newblock {\emph{\JournalTitle{Nano Letters}}} \textbf{\bibinfo{volume}{10}},
	\bibinfo{pages}{1271--1275}, \doiprefix\url{10.1021/nl903868w}
	(\bibinfo{year}{2010}).
	
	\bibitem{mak_atomically_2010}
	\bibinfo{author}{Mak, K.~F.}, \bibinfo{author}{Lee, C.}, \bibinfo{author}{Hone,
		J.}, \bibinfo{author}{Shan, J.} \& \bibinfo{author}{Heinz, T.~F.}
	\newblock \bibinfo{journal}{\bibinfo{title}{Atomically {Thin} {MoS}\_\{2\}: {A}
			{New} {Direct}-{Gap} {Semiconductor}}}.
	\newblock {\emph{\JournalTitle{Physical Review Letters}}}
	\textbf{\bibinfo{volume}{105}}, \bibinfo{pages}{136805},
	\doiprefix\url{10.1103/PhysRevLett.105.136805} (\bibinfo{year}{2010}).
	
	\bibitem{zhao_evolution_2013}
	\bibinfo{author}{Zhao, W.} \emph{et~al.}
	\newblock \bibinfo{journal}{\bibinfo{title}{Evolution of {Electronic}
			{Structure} in {Atomically} {Thin} {Sheets} of {WS2} and {WSe2}}}.
	\newblock {\emph{\JournalTitle{ACS Nano}}} \textbf{\bibinfo{volume}{7}},
	\bibinfo{pages}{791--797}, \doiprefix\url{10.1021/nn305275h}
	(\bibinfo{year}{2013}).
	
	\bibitem{stroscio_scanning_1993}
	\bibinfo{author}{Stroscio, J.~A.} \& \bibinfo{author}{Kaiser, W.~J.}
	\newblock \emph{\bibinfo{title}{Scanning {Tunneling} {Microscopy}}}
	(\bibinfo{publisher}{Elsevier}, \bibinfo{year}{1993}).
	
	\bibitem{feenstra_tunneling_1987}
	\bibinfo{author}{Feenstra, R.~M.} \& \bibinfo{author}{Stroscio, J.~A.}
	\newblock \bibinfo{journal}{\bibinfo{title}{Tunneling spectroscopy of the
			{GaAs}(110) surface}}.
	\newblock {\emph{\JournalTitle{Journal of Vacuum Science \& Technology B:
				Microelectronics Processing and Phenomena}}} \textbf{\bibinfo{volume}{5}},
	\bibinfo{pages}{923--929}, \doiprefix\url{10.1116/1.583691}
	(\bibinfo{year}{1987}).
	\newblock \bibinfo{note}{Publisher: American Institute of Physics}.
	
	\bibitem{park_direct_2018}
	\bibinfo{author}{Park, S.} \emph{et~al.}
	\newblock \bibinfo{journal}{\bibinfo{title}{Direct determination of monolayer
			{MoS} 2 and {WSe} 2 exciton binding energies on insulating and metallic
			substrates}}.
	\newblock {\emph{\JournalTitle{2D Materials}}} \textbf{\bibinfo{volume}{5}},
	\bibinfo{pages}{025003}, \doiprefix\url{10.1088/2053-1583/aaa4ca}
	(\bibinfo{year}{2018}).
	\newblock \bibinfo{note}{Publisher: IOP Publishing}.
	
	\bibitem{wilson_determination_2017}
	\bibinfo{author}{Wilson, N.~R.} \emph{et~al.}
	\newblock \bibinfo{journal}{\bibinfo{title}{Determination of band offsets,
			hybridization, and exciton binding in {2D} semiconductor heterostructures}}.
	\newblock {\emph{\JournalTitle{Science Advances}}}
	\textbf{\bibinfo{volume}{3}}, \bibinfo{pages}{e1601832},
	\doiprefix\url{10.1126/sciadv.1601832} (\bibinfo{year}{2017}).
	\newblock \bibinfo{note}{Publisher: American Association for the Advancement of
		Science Section: Research Article}.
	
	\bibitem{cucchi_microfocus_2019}
	\bibinfo{author}{Cucchi, I.} \emph{et~al.}
	\newblock \bibinfo{journal}{\bibinfo{title}{Microfocus
			{Laser}–{Angle}-{Resolved} {Photoemission} on {Encapsulated} {Mono}-,
			{Bi}-, and {Few}-{Layer} {1T}'-{WTe2}}}.
	\newblock {\emph{\JournalTitle{Nano Letters}}} \textbf{\bibinfo{volume}{19}},
	\bibinfo{pages}{554--560}, \doiprefix\url{10.1021/acs.nanolett.8b04534}
	(\bibinfo{year}{2019}).
	\newblock \bibinfo{note}{Publisher: American Chemical Society}.
	
	\bibitem{hamer_indirect_2019}
	\bibinfo{author}{Hamer, M.~J.} \emph{et~al.}
	\newblock \bibinfo{journal}{\bibinfo{title}{Indirect to {Direct} {Gap}
			{Crossover} in {Two}-{Dimensional} {InSe} {Revealed} by {Angle}-{Resolved}
			{Photoemission} {Spectroscopy}}}.
	\newblock {\emph{\JournalTitle{ACS Nano}}} \textbf{\bibinfo{volume}{13}},
	\bibinfo{pages}{2136--2142}, \doiprefix\url{10.1021/acsnano.8b08726}
	(\bibinfo{year}{2019}).
	\newblock \bibinfo{note}{Publisher: American Chemical Society}.
	
	\bibitem{riley_negative_2015}
	\bibinfo{author}{Riley, J.~M.} \emph{et~al.}
	\newblock \bibinfo{journal}{\bibinfo{title}{Negative electronic compressibility
			and tunable spin splitting in {WSe2}}}.
	\newblock {\emph{\JournalTitle{Nature Nanotechnology}}}
	\textbf{\bibinfo{volume}{10}}, \bibinfo{pages}{1043--1047},
	\doiprefix\url{10.1038/nnano.2015.217} (\bibinfo{year}{2015}).
	
	\bibitem{kim_possible_2017}
	\bibinfo{author}{Kim, B.~S.} \emph{et~al.}
	\newblock \bibinfo{journal}{\bibinfo{title}{Possible electric field induced
			indirect to direct band gap transition in {MoSe2}}}.
	\newblock {\emph{\JournalTitle{Scientific Reports}}}
	\textbf{\bibinfo{volume}{7}}, \bibinfo{pages}{5206},
	\doiprefix\url{10.1038/s41598-017-05613-5} (\bibinfo{year}{2017}).
	
	\bibitem{kang_universal_2017}
	\bibinfo{author}{Kang, M.} \emph{et~al.}
	\newblock \bibinfo{journal}{\bibinfo{title}{Universal {Mechanism} of
			{Band}-{Gap} {Engineering} in {Transition}-{Metal} {Dichalcogenides}}}.
	\newblock {\emph{\JournalTitle{Nano Letters}}} \textbf{\bibinfo{volume}{17}},
	\bibinfo{pages}{1610--1615}, \doiprefix\url{10.1021/acs.nanolett.6b04775}
	(\bibinfo{year}{2017}).
	
	\bibitem{katoch_giant_2018}
	\bibinfo{author}{Katoch, J.} \emph{et~al.}
	\newblock \bibinfo{journal}{\bibinfo{title}{Giant spin-splitting and gap
			renormalization driven by trions in single-layer {WS} 2 /h-{BN}
			heterostructures}}.
	\newblock {\emph{\JournalTitle{Nature Physics}}} \textbf{\bibinfo{volume}{14}},
	\bibinfo{pages}{355--359}, \doiprefix\url{10.1038/s41567-017-0033-4}
	(\bibinfo{year}{2018}).
	\newblock \bibinfo{note}{Number: 4 Publisher: Nature Publishing Group}.
	
	\bibitem{nguyen_visualizing_2019}
	\bibinfo{author}{Nguyen, P.~V.} \emph{et~al.}
	\newblock \bibinfo{journal}{\bibinfo{title}{Visualizing electrostatic gating
			effects in two-dimensional heterostructures}}.
	\newblock {\emph{\JournalTitle{Nature}}} \textbf{\bibinfo{volume}{572}},
	\bibinfo{pages}{220--223}, \doiprefix\url{10.1038/s41586-019-1402-1}
	(\bibinfo{year}{2019}).
	\newblock \bibinfo{note}{Number: 7768 Publisher: Nature Publishing Group}.
	
	\bibitem{kouwenhoven_electron_1997}
	\bibinfo{author}{Kouwenhoven, L.~P.} \emph{et~al.}
	\newblock \bibinfo{title}{Electron {Transport} in {Quantum} {Dots}}.
	\newblock In \bibinfo{editor}{Sohn, L.~L.}, \bibinfo{editor}{Kouwenhoven,
		L.~P.} \& \bibinfo{editor}{Schön, G.} (eds.)
	\emph{\bibinfo{booktitle}{Mesoscopic {Electron} {Transport}}}, {NATO} {ASI}
	{Series}, \bibinfo{pages}{105--214},
	\doiprefix\url{10.1007/978-94-015-8839-3_4} (\bibinfo{publisher}{Springer
		Netherlands}, \bibinfo{address}{Dordrecht}, \bibinfo{year}{1997}).
	
	\bibitem{shimotani_continuous_2014}
	\bibinfo{author}{Shimotani, H.} \emph{et~al.}
	\newblock \bibinfo{journal}{\bibinfo{title}{Continuous {Band}-{Filling}
			{Control} and {One}-{Dimensional} {Transport} in {Metallic} and
			{Semiconducting} {Carbon} {Nanotube} {Tangled} {Films}}}.
	\newblock {\emph{\JournalTitle{Advanced Functional Materials}}}
	\textbf{\bibinfo{volume}{24}}, \bibinfo{pages}{3305--3311},
	\doiprefix\url{10.1002/adfm.201303566} (\bibinfo{year}{2014}).
	\newblock \bibinfo{note}{\_eprint:
		https://onlinelibrary.wiley.com/doi/pdf/10.1002/adfm.201303566}.
	
	\bibitem{mounet_two-dimensional_2018}
	\bibinfo{author}{Mounet, N.} \emph{et~al.}
	\newblock \bibinfo{journal}{\bibinfo{title}{Two-dimensional materials from
			high-throughput computational exfoliation of experimentally known
			compounds}}.
	\newblock {\emph{\JournalTitle{Nature Nanotechnology}}}
	\textbf{\bibinfo{volume}{13}}, \bibinfo{pages}{246--252},
	\doiprefix\url{10.1038/s41565-017-0035-5} (\bibinfo{year}{2018}).
	\newblock \bibinfo{note}{Number: 3 Publisher: Nature Publishing Group}.
	
	\bibitem{podzorov_high-mobility_2004}
	\bibinfo{author}{Podzorov, V.}, \bibinfo{author}{Gershenson, M.~E.},
	\bibinfo{author}{Kloc, C.}, \bibinfo{author}{Zeis, R.} \&
	\bibinfo{author}{Bucher, E.}
	\newblock \bibinfo{journal}{\bibinfo{title}{High-mobility field-effect
			transistors based on transition metal dichalcogenides}}.
	\newblock {\emph{\JournalTitle{Applied Physics Letters}}}
	\textbf{\bibinfo{volume}{84}}, \bibinfo{pages}{3301--3303},
	\doiprefix\url{10.1063/1.1723695} (\bibinfo{year}{2004}).
	\newblock \bibinfo{note}{Publisher: American Institute of Physics}.
	
	\bibitem{ramasubramaniam_tunable_2011}
	\bibinfo{author}{Ramasubramaniam, A.}, \bibinfo{author}{Naveh, D.} \&
	\bibinfo{author}{Towe, E.}
	\newblock \bibinfo{journal}{\bibinfo{title}{Tunable band gaps in bilayer
			transition-metal dichalcogenides}}.
	\newblock {\emph{\JournalTitle{Physical Review B}}}
	\textbf{\bibinfo{volume}{84}}, \bibinfo{pages}{205325},
	\doiprefix\url{10.1103/PhysRevB.84.205325} (\bibinfo{year}{2011}).
	
	\bibitem{brumme_first-principles_2015}
	\bibinfo{author}{Brumme, T.}, \bibinfo{author}{Calandra, M.} \&
	\bibinfo{author}{Mauri, F.}
	\newblock \bibinfo{journal}{\bibinfo{title}{First-principles theory of
			field-effect doping in transition-metal dichalcogenides: {Structural}
			properties, electronic structure, {Hall} coefficient, and electrical
			conductivity}}.
	\newblock {\emph{\JournalTitle{Physical Review B}}}
	\textbf{\bibinfo{volume}{91}}, \bibinfo{pages}{155436},
	\doiprefix\url{10.1103/PhysRevB.91.155436} (\bibinfo{year}{2015}).
	\newblock \bibinfo{note}{Publisher: American Physical Society}.
	
	\bibitem{shimotani_gate_2006}
	\bibinfo{author}{Shimotani, H.} \emph{et~al.}
	\newblock \bibinfo{journal}{\bibinfo{title}{Gate capacitance in electrochemical
			transistor of single-walled carbon nanotube}}.
	\newblock {\emph{\JournalTitle{Applied Physics Letters}}}
	\textbf{\bibinfo{volume}{88}}, \bibinfo{pages}{073104},
	\doiprefix\url{10.1063/1.2173626} (\bibinfo{year}{2006}).
	\newblock \bibinfo{note}{Publisher: American Institute of Physics}.
	
	\bibitem{bard_electrochemical_nodate}
	\bibinfo{author}{Bard, A.~J.} \& \bibinfo{author}{Faulkner, L.~R.}
	\newblock \emph{\bibinfo{title}{Electrochemical {Methods}}}.
	
	\bibitem{ponomarev_hole_2018}
	\bibinfo{author}{Ponomarev, E.} \emph{et~al.}
	\newblock \bibinfo{journal}{\bibinfo{title}{Hole {Transport} in {Exfoliated}
			{Monolayer} {MoS2}}}.
	\newblock {\emph{\JournalTitle{ACS Nano}}} \textbf{\bibinfo{volume}{12}},
	\bibinfo{pages}{2669--2676}, \doiprefix\url{10.1021/acsnano.7b08831}
	(\bibinfo{year}{2018}).
	\newblock \bibinfo{note}{Publisher: American Chemical Society}.
	
	\bibitem{vasu_opening_2018}
	\bibinfo{author}{Vasu, K.~S.} \emph{et~al.}
	\newblock \bibinfo{journal}{\bibinfo{title}{Opening of large band gaps in
			metallic carbon nanotubes by mannose-functionalized dendrimers: experiments
			and theory}}.
	\newblock {\emph{\JournalTitle{Journal of Materials Chemistry C}}}
	\textbf{\bibinfo{volume}{6}}, \bibinfo{pages}{6483--6488},
	\doiprefix\url{10.1039/C8TC01269E} (\bibinfo{year}{2018}).
	\newblock \bibinfo{note}{Publisher: The Royal Society of Chemistry}.
	
	\bibitem{xu_reconfigurable_2015}
	\bibinfo{author}{Xu, H.}, \bibinfo{author}{Fathipour, S.},
	\bibinfo{author}{Kinder, E.~W.}, \bibinfo{author}{Seabaugh, A.~C.} \&
	\bibinfo{author}{Fullerton-Shirey, S.~K.}
	\newblock \bibinfo{journal}{\bibinfo{title}{Reconfigurable {Ion} {Gating} of
			{2H}-{MoTe2} {Field}-{Effect} {Transistors} {Using} {Poly}(ethylene
			oxide)-{CsClO4} {Solid} {Polymer} {Electrolyte}}}.
	\newblock {\emph{\JournalTitle{ACS Nano}}} \textbf{\bibinfo{volume}{9}},
	\bibinfo{pages}{4900--4910}, \doiprefix\url{10.1021/nn506521p}
	(\bibinfo{year}{2015}).
	\newblock \bibinfo{note}{Publisher: American Chemical Society}.
	
	\bibitem{prakash_bandgap_2017}
	\bibinfo{author}{Prakash, A.} \& \bibinfo{author}{Appenzeller, J.}
	\newblock \bibinfo{journal}{\bibinfo{title}{Bandgap {Extraction} and {Device}
			{Analysis} of {Ionic} {Liquid} {Gated} {WSe2} {Schottky} {Barrier}
			{Transistors}}}.
	\newblock {\emph{\JournalTitle{ACS Nano}}} \textbf{\bibinfo{volume}{11}},
	\bibinfo{pages}{1626--1632}, \doiprefix\url{10.1021/acsnano.6b07360}
	(\bibinfo{year}{2017}).
	\newblock \bibinfo{note}{Publisher: American Chemical Society}.
	
	\bibitem{yomogida_ambipolar_2012}
	\bibinfo{author}{Yomogida, Y.} \emph{et~al.}
	\newblock \bibinfo{journal}{\bibinfo{title}{Ambipolar {Organic}
			{Single}-{Crystal} {Transistors} {Based} on {Ion} {Gels}}}.
	\newblock {\emph{\JournalTitle{Advanced Materials}}}
	\textbf{\bibinfo{volume}{24}}, \bibinfo{pages}{4392--4397},
	\doiprefix\url{10.1002/adma.201200655} (\bibinfo{year}{2012}).
	\newblock \bibinfo{note}{\_eprint:
		https://onlinelibrary.wiley.com/doi/pdf/10.1002/adma.201200655}.
	
	\bibitem{rosner_two-dimensional_2016}
	\bibinfo{author}{Rösner, M.} \emph{et~al.}
	\newblock \bibinfo{journal}{\bibinfo{title}{Two-{Dimensional} {Heterojunctions}
			from {Nonlocal} {Manipulations} of the {Interactions}}}.
	\newblock {\emph{\JournalTitle{Nano Letters}}} \textbf{\bibinfo{volume}{16}},
	\bibinfo{pages}{2322--2327}, \doiprefix\url{10.1021/acs.nanolett.5b05009}
	(\bibinfo{year}{2016}).
	\newblock \bibinfo{note}{Publisher: American Chemical Society}.
	
	\bibitem{cho_environmentally_2018}
	\bibinfo{author}{Cho, Y.} \& \bibinfo{author}{Berkelbach, T.~C.}
	\newblock \bibinfo{journal}{\bibinfo{title}{Environmentally sensitive theory of
			electronic and optical transitions in atomically thin semiconductors}}.
	\newblock {\emph{\JournalTitle{Physical Review B}}}
	\textbf{\bibinfo{volume}{97}}, \bibinfo{pages}{041409},
	\doiprefix\url{10.1103/PhysRevB.97.041409} (\bibinfo{year}{2018}).
	\newblock \bibinfo{note}{Publisher: American Physical Society}.
	
	\bibitem{waldecker_rigid_2019}
	\bibinfo{author}{Waldecker, L.} \emph{et~al.}
	\newblock \bibinfo{journal}{\bibinfo{title}{Rigid {Band} {Shifts} in
			{Two}-{Dimensional} {Semiconductors} through {External} {Dielectric}
			{Screening}}}.
	\newblock {\emph{\JournalTitle{Physical Review Letters}}}
	\textbf{\bibinfo{volume}{123}}, \bibinfo{pages}{206403},
	\doiprefix\url{10.1103/PhysRevLett.123.206403} (\bibinfo{year}{2019}).
	\newblock \bibinfo{note}{Publisher: American Physical Society}.
	
	\bibitem{geim_van_2013}
	\bibinfo{author}{Geim, A.~K.} \& \bibinfo{author}{Grigorieva, I.~V.}
	\newblock \bibinfo{journal}{\bibinfo{title}{Van der {Waals} heterostructures}}.
	\newblock {\emph{\JournalTitle{Nature}}} \textbf{\bibinfo{volume}{499}},
	\bibinfo{pages}{419--425}, \doiprefix\url{10.1038/nature12385}
	(\bibinfo{year}{2013}).
	\newblock \bibinfo{note}{Number: 7459 Publisher: Nature Publishing Group}.
	
	\bibitem{novoselov_2d_2016}
	\bibinfo{author}{Novoselov, K.~S.}, \bibinfo{author}{Mishchenko, A.},
	\bibinfo{author}{Carvalho, A.} \& \bibinfo{author}{Neto, A. H.~C.}
	\newblock \bibinfo{journal}{\bibinfo{title}{{2D} materials and van der {Waals}
			heterostructures}}.
	\newblock {\emph{\JournalTitle{Science}}} \textbf{\bibinfo{volume}{353}},
	\doiprefix\url{10.1126/science.aac9439} (\bibinfo{year}{2016}).
	\newblock \bibinfo{note}{Publisher: American Association for the Advancement of
		Science Section: Review}.
	
	\bibitem{rivera_observation_2015}
	\bibinfo{author}{Rivera, P.} \emph{et~al.}
	\newblock \bibinfo{journal}{\bibinfo{title}{Observation of long-lived
			interlayer excitons in monolayer {MoSe2}–{WSe2} heterostructures}}.
	\newblock {\emph{\JournalTitle{Nature Communications}}}
	\textbf{\bibinfo{volume}{6}}, \bibinfo{pages}{6242},
	\doiprefix\url{10.1038/ncomms7242} (\bibinfo{year}{2015}).
	
	\bibitem{aretouli_epitaxial_2016}
	\bibinfo{author}{Aretouli, K.~E.} \emph{et~al.}
	\newblock \bibinfo{journal}{\bibinfo{title}{Epitaxial {2D} {SnSe2}/ {2D} {WSe2}
			van der {Waals} {Heterostructures}}}.
	\newblock {\emph{\JournalTitle{ACS Applied Materials \& Interfaces}}}
	\textbf{\bibinfo{volume}{8}}, \bibinfo{pages}{23222--23229},
	\doiprefix\url{10.1021/acsami.6b02933} (\bibinfo{year}{2016}).
	\newblock \bibinfo{note}{Publisher: American Chemical Society}.
	
	\bibitem{yan_tunable_2017}
	\bibinfo{author}{Yan, X.} \emph{et~al.}
	\newblock \bibinfo{journal}{\bibinfo{title}{Tunable {SnSe2}/{WSe2}
			{Heterostructure} {Tunneling} {Field} {Effect} {Transistor}}}.
	\newblock {\emph{\JournalTitle{Small}}} \textbf{\bibinfo{volume}{13}},
	\bibinfo{pages}{1701478}, \doiprefix\url{10.1002/smll.201701478}
	(\bibinfo{year}{2017}).
	\newblock \bibinfo{note}{\_eprint:
		https://onlinelibrary.wiley.com/doi/pdf/10.1002/smll.201701478}.
	
	\bibitem{gibertini_magnetic_2019}
	\bibinfo{author}{Gibertini, M.}, \bibinfo{author}{Koperski, M.},
	\bibinfo{author}{Morpurgo, A.~F.} \& \bibinfo{author}{Novoselov, K.~S.}
	\newblock \bibinfo{journal}{\bibinfo{title}{Magnetic {2D} materials and
			heterostructures}}.
	\newblock {\emph{\JournalTitle{Nature Nanotechnology}}}
	\textbf{\bibinfo{volume}{14}}, \bibinfo{pages}{408--419},
	\doiprefix\url{10.1038/s41565-019-0438-6} (\bibinfo{year}{2019}).
	\newblock \bibinfo{note}{Number: 5 Publisher: Nature Publishing Group}.
	
	\bibitem{bandurin_high_2017}
	\bibinfo{author}{Bandurin, D.~A.} \emph{et~al.}
	\newblock \bibinfo{journal}{\bibinfo{title}{High electron mobility, quantum
			{Hall} effect and anomalous optical response in atomically thin {InSe}}}.
	\newblock {\emph{\JournalTitle{Nature Nanotechnology}}}
	\textbf{\bibinfo{volume}{12}}, \bibinfo{pages}{223--227},
	\doiprefix\url{10.1038/nnano.2016.242} (\bibinfo{year}{2017}).
	
	\bibitem{guo_band_2017}
	\bibinfo{author}{Guo, Y.} \& \bibinfo{author}{Robertson, J.}
	\newblock \bibinfo{journal}{\bibinfo{title}{Band structure, band offsets,
			substitutional doping, and {Schottky} barriers of bulk and monolayer
			{InSe}}}.
	\newblock {\emph{\JournalTitle{Physical Review Materials}}}
	\textbf{\bibinfo{volume}{1}}, \bibinfo{pages}{044004},
	\doiprefix\url{10.1103/PhysRevMaterials.1.044004} (\bibinfo{year}{2017}).
	\newblock \bibinfo{note}{Publisher: American Physical Society}.
	
	\bibitem{ye_accessing_2011}
	\bibinfo{author}{Ye, J.} \emph{et~al.}
	\newblock \bibinfo{journal}{\bibinfo{title}{Accessing the transport properties
			of graphene and its multilayers at high carrier density}}.
	\newblock {\emph{\JournalTitle{Proceedings of the National Academy of
				Sciences}}} \textbf{\bibinfo{volume}{108}}, \bibinfo{pages}{13002--13006},
	\doiprefix\url{10.1073/pnas.1018388108} (\bibinfo{year}{2011}).
	
	\bibitem{gonnelli_weak_2017}
	\bibinfo{author}{Gonnelli, R.~S.} \emph{et~al.}
	\newblock \bibinfo{journal}{\bibinfo{title}{Weak localization in
			electric-double-layer gated few-layer graphene}}.
	\newblock {\emph{\JournalTitle{2D Materials}}} \textbf{\bibinfo{volume}{4}},
	\bibinfo{pages}{035006}, \doiprefix\url{10.1088/2053-1583/aa5afe}
	(\bibinfo{year}{2017}).
	\newblock \bibinfo{note}{Publisher: IOP Publishing}.
	
\end{thebibliography}

\section*{Acknowledgements}
We acknowledge Alexandre Ferreira for his continuous and precious technical support over all years. We acknowledge the scientific contributions of several former and current group members who have been actively involved in the development of the technique: Daniele Braga, Sanghyun Jo, Marc Philippi, Aditya Reddy, and Haijing Zhang. Financial support from the Swiss National Science Foundation and from the EU Graphene Flagship Project is also gratefully acknowledged.


\section*{Competing interests}
The authors declare no competing interests.


\newpage

\section*{Box 1}
\begin{figure}[ht]
	\centering
	\includegraphics[width=.9\textwidth]{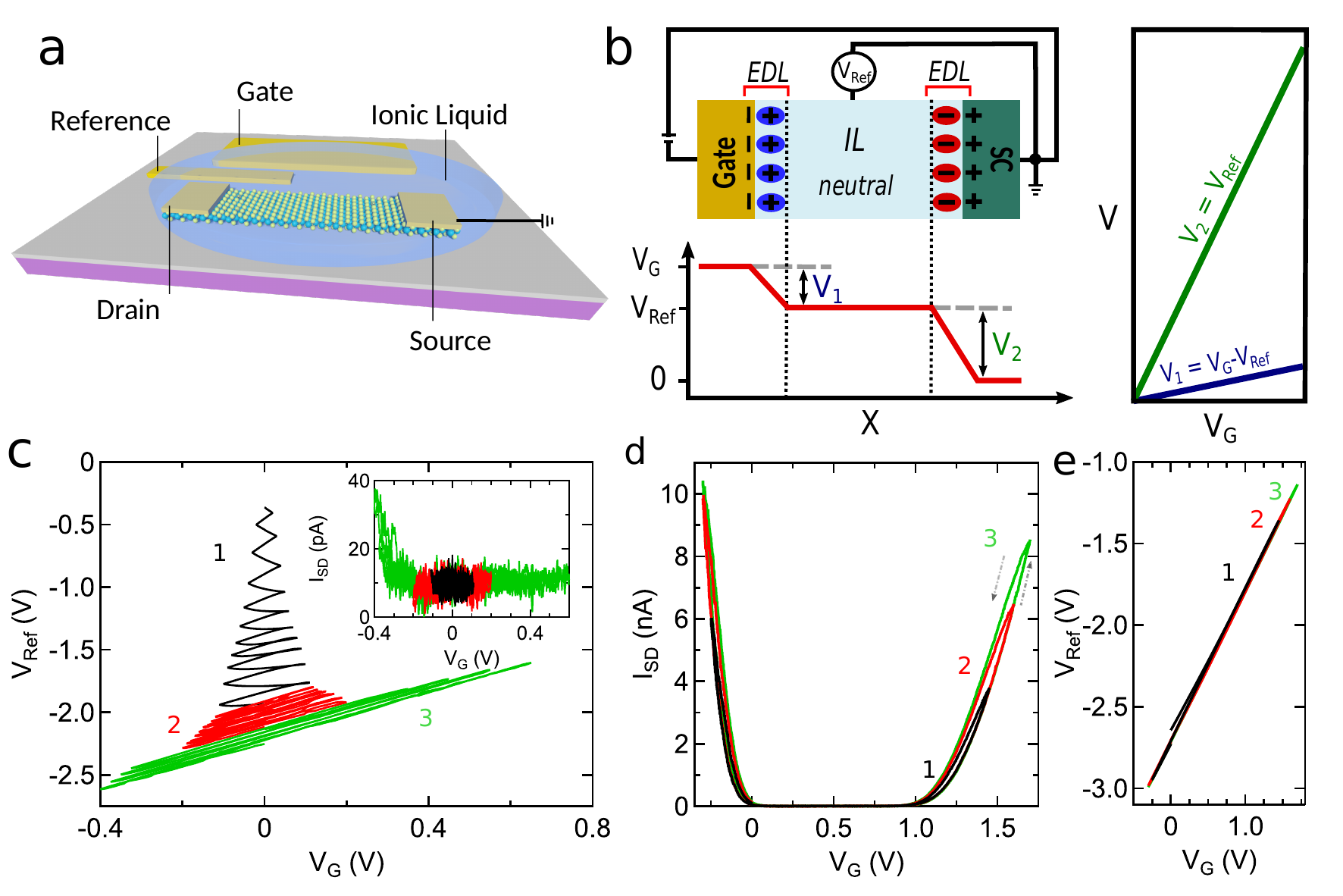}
	\caption{
	}
	\label{fig:1}
\end{figure}
In an ion-gated field-effect transistor, a semiconductor with source and drain electrodes is in direct contact with an electrolyte containing a very high density of mobile charged ions; the same electrolyte is also in contact with the gate and the reference electrode (see Fig.~\figref{fig:1}{a}). The voltage applied between gate and drain causes the accumulation of oppositely charged ions at the two interfaces with the electrolyte, resulting in the formation of so-called double layers (typically $\approx$~1~nm thick or less) that screen the electric field\cite{fujimoto_electric-double-layer_2013}. The body of the electrolyte is then at a uniform potential, whose value is determined by how the applied voltage is capacitively partitioned between the gate/electrolyte and the electrolyte/semiconductor interfaces (Fig.~\figref{fig:1}{b}). The gate electrode is intentionally fabricated to have a very large surface area, resulting in a correspondingly large capacitance, to  make the voltage drop at the gate/electrolyte interface ($V_1$) negligible. Under this condition, the applied gate voltage drops entirely at the electrolyte/semiconductor interface, and the gate voltage coincides with the potential of the electrolyte measured by the reference electrode, \ie\ \vref~=~$V_2$~=~$V_G$ (more precisely, $\Delta\vref = \Delta V_G$, since any difference in work function between the gate and the gated material gives a non-zero built-in voltage difference already present at $V_G$~=~0~V).\\

In current experiments on two-dimensional semiconductors, the electrolyte is normally an ionic liquid. Devices are operated in a vacuum chamber ( $p\approx 10^{-6}$~mbar) and pumped overnight, to remove humidity and oxygen that upon biasing can cause unwanted chemical reactions and device degradation. As illustrated in Fig.~\figref{fig:1}{c}, device operation starts by sweeping slowly the gate voltage over a small range that is gradually increased, to eventually reach the full range of interest (typically between approximately -3~V and +3~V, depending on the material to be gated). Throughout this process, the voltage measured at the reference potential, as well as other transistor characteristics, are commonly observed to drift before eventually stabilizing. The drift is attributed to the desorption of adsorbates present at the surface of the gate and of the semiconductor, which modify the work function as well as the interfacial capacitance, and that can affect the electronic properties of the semiconductor in different ways. If the gate voltage is increased too rapidly, these same adsorbates can react chemically and cause irreversible degradation. Increasing the gate voltage range only very slowly is therefore essential.\\ 

When the device has stabilized, reproducible curves are measured upon repeatedly sweeping the gate voltage and spectroscopic measurements can be performed  (see Fig.~\figref{fig:1}{d} and \figref{fig:1}{e}). The spectroscopic capabilities of ionic liquid gated transistors come from the relation $e \Delta V_G = \Delta \mu $ (see discussion in the main text). To ensure the validity of this condition experimentally it is key that virtually no current flows through the electrolyte, since any non-negligible current causes the electrolyte not to be equipotential, and prevents the analysis of the voltages in terms of a simple capacitive model. That is why it is extremely important to constantly monitor the gate leakage current when operating the devices, and to maintain it negligibly small by sweeping the gate voltage sufficiently slowly (the displacement current increases linearly with  sweeping rate). A rapid increase of leakage current upon increasing $V_G$ may also be indicative of chemical reactions that can cause device degradation, and should therefore be avoided.\\

\FloatBarrier
\section*{Box 2}
\begin{figure}[ht]
	\centering
	\includegraphics[width=.9\textwidth]{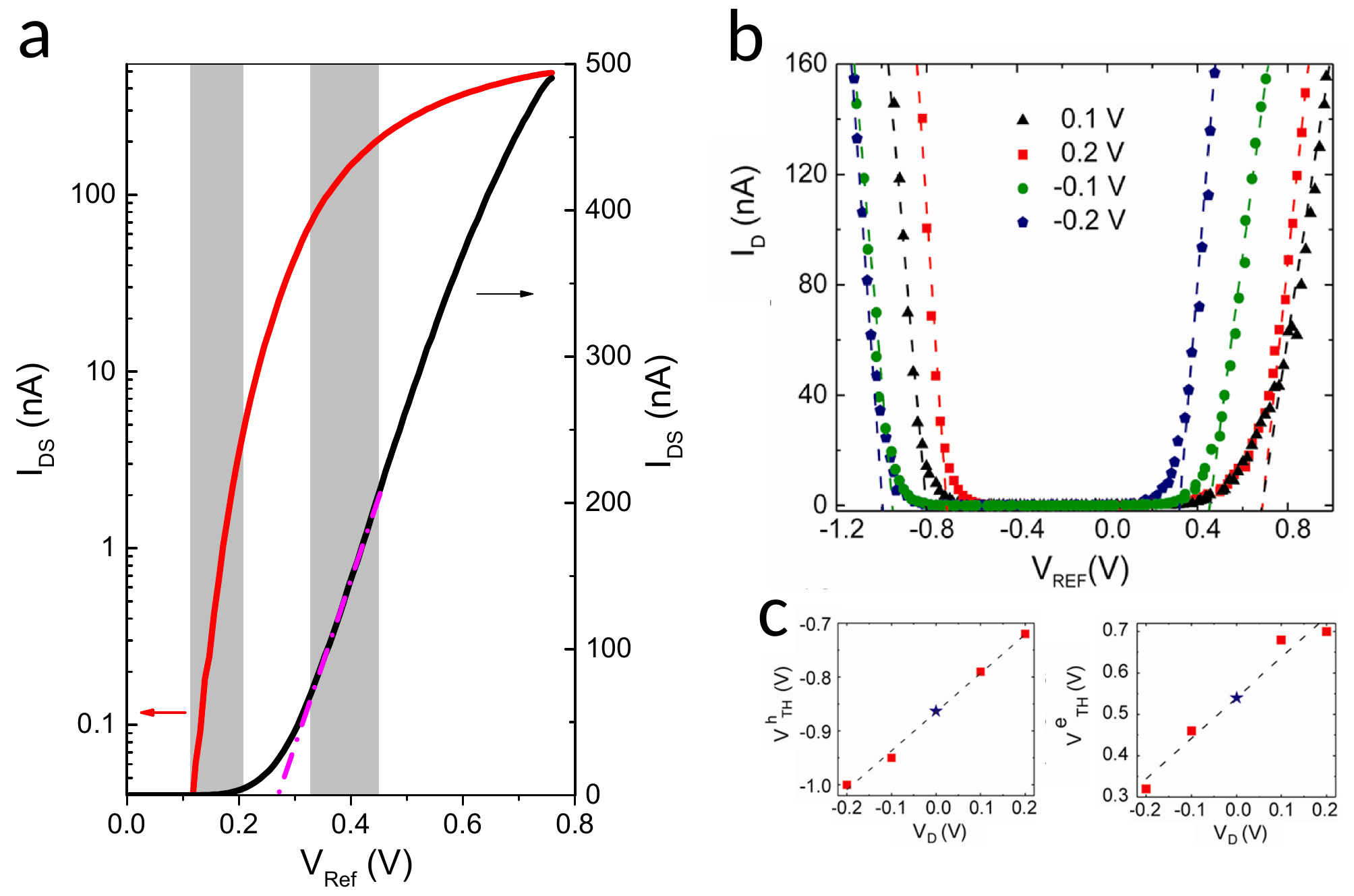}
	\caption{}
	\label{box:2}
\end{figure}

Ionic gate spectroscopy relies on the ability to extract precise and reliable values of threshold voltage from the transistor characteristics. By definition, the threshold voltage $V_{th}$ is the value of $V_G$  obtained by extrapolating to zero the source-drain $I_{SD}(V_G)$ curve  measured in the linear regime. To perform the extrapolation properly, the entire gate voltage range considered must be out of the sub-threshold region, in which $I_{SD}$ increases exponentially on $V_G$. This can be easily ensured by comparing the plots of $I_{SD}$-vs-$V_G$ in logarithmic and linear scale (see Fig.~\ref{box:2}{\bfseries a}). The $V_G$ range also does not have to extend up to too large values, which would result in a sizable  perpendicular electric field that can affect the band structure of the material. It is therefore important to identify a sufficiently large interval of $V_G$ enabling a precise extrapolation, but not too large. 

As there is not a well-defined procedure to determine the $V_G$ interval,  it is important to check the band gap values  extracted in different ways.  Comparing the $V_{th}$ values extracted from different intervals of $V_G$ is a simple but effective possibility. Alternatively, the threshold voltage can be extracted from $I_{SD}(V_G)$ curves measured with  different applied source-drain bias $V_{SD}$ (see Fig.~\ref{box:2}{\bfseries b}; Adapted from Ref.~\cite{braga_quantitative_2012}). A variation in \vsd\ causes a linear shift in threshold voltage: taking the $V_{th}$  value obtained  by extrapolating to $V_{SD}=0$ V (see Fig.~\ref{box:2}{\bfseries c}; Adapted from Ref.~\cite{braga_quantitative_2012}) increases the precision, because uncertainties present in the measurements at different \vsd\ compensate\cite{braga_quantitative_2012}. Finally, it is always essential to measure multiple devices and compare the results. 

We have followed these procedures in multiple occasions and found that for the materials that we have investigated with  a band gap  between 1 and 2 eV, the uncertainty is typically less than $\pm 5$ \%. In some cases,  this precision could also be achieved in devices  exhibiting non-idealities in their transfer curves, such as hysteresis (with the $I_{SD}(V_G)$ curve traced upon sweeping $V_G$ from positive to negative voltage, differing from that measured upon sweeping $V_G$ in the different direction). If --despite the hysteresis-- multiple measurements lead to exactly the same $I_{SD}(V_G)$ curve, the correct value of the band gap could be extracted as  the difference of threshold voltages for electron and hole transport measured for a same sweep direction of  $V_G$ (see Ref.~\cite{gutierrez-lezama_electroluminescence_2016} for details). 

\end{document}